# Mechanism of Na-ionic conduction in the high efficient layered battery material $Na_2Mn_3O_7$

*Bikash Saha,*[1,2] *A. K. Bera*[1,*] *and S. M. Yusuf*[1,2,*]

1Solid State Physics Division, Bhabha Atomic Research Centre, Mumbai 400085, India
2Homi Bhabha National Institute, Anushaktinagar, Mumbai 400094, India
* E-mails: akbera@barc.gov.in, smyusuf@barc.gov.in

**ABSTRACT**

The ionic conduction properties of the technologically important two-dimensional (2D) layered battery material $Na_2Mn_3O_7$, with exceptional small-voltage hysteresis between charge and discharge curves, have been investigated as a function of temperature and frequency by an impedance spectroscopy. The detailed analyses of the impedance data in the form of dc-conductivity, ac-conductivity, electrical modulus, dielectric constant and complex polarizability reveal a long-range Na-ionic conductivity with negligible contribution from a local dipole relaxation. A significant enhancement (~$10^4$ times) of the Na-ion conductivity has been found with the increasing temperature from 353 K to 713 K. The temperature dependent conductivity reveals thermally activated conduction process with activation energies of 0.161 and 0.377 eV over the two temperature regions of 383– 518 K and 518– 713 K, respectively. AC conductivity study reveals a long-range hopping process for the conduction of charge carriers with a sharp increase of the hopping range at 518 K. With the increasing frequency, the activation energies decrease for both the temperature regions. The scaling study of the ac-conductivity reveals that the frequency-activated conductivity (above $\nu_C = 10^4$ Hz at 353 K) is mainly controlled by the critical frequency ($\nu_C$) that increases with the increasing temperature. Our results reveal that the thermally activated Na-ion conduction in the present 2D layered compound $Na_2Mn_3O_7$ occurs predominantly by a correlated barrier hopping process. Besides, a correlation between ionic conduction and crystal structure has been established by x-ray and neutron diffraction study. We have further shown that the conductivity of $Na_2Mn_3O_7$ can be enhanced by reduction of the stacking faults in the crystal structure. The present comprehensive study facilitates the understanding of the microscopic ionic conduction mechanism in the highly efficient 2D layered battery material $Na_2Mn_3O_7$ having high energy storage capacity and high structural stability, paving way for the discovery of 2D materials for functional battery applications.

## I. INTRODUCTION

The demand for advanced energy storage systems is increasing worldwide to store and optimised-utilization of renewable energy.[1-2] The requirement for inexpensive and environment friendly energy storage devices has led to the possibility of Na ion-based battery materials as a substitute to presently used Li ion battery materials. Search for novel Na-ion based battery electrode materials considering the key factors, such as fast Na ion extraction/insertion, energy density, cost, toxicity, reversible capacity, and cyclic stability is a subject of research interest world wise. It is well known that a suitable electrode must have good ionic conductivity in order to exhibit high efficient battery performance.[3-5] The parameter conductivity plays a significant role in intercalation/deintercalation -based energy storage devices where charge transport of $Na^+$ ions takes place within the electrode materials. It is often found that a considerable drop in battery performance at high charge/discharge rates is due to a slow movement of $Na^+$ ions through the material. Various approach via structure modification, introducing defects, selective doping with transition metal ions, etc. have been put forward to improve the ionic conductivity.[6] The electrical response is dictated by the microstructural network. The understanding of the microscopic conduction mechanism is a key requirement for the design of high-performance battery materials, and advancement of the battery technology.

Na-ion based compounds with two-dimensional (2D) layered crystal structure, especially layered transition metal oxides having intermediate layers of Na-ions alone, are of special interest in this matter due to their high ionic conductivity, high energy density, low dissipation energy, and anisotropic nature arising from the layered structure.[7-11] For these 2D materials, the absence of other ions within the Na layers i.e., the absence of the scattering centers due to other ions leads to the high ionic conductions. Moreover, layered oxide materials have a degree of freedom in their crystal structure to expand the interstitial space between strongly bonded 2D layers providing better intercalation/deintercalation of Na ions in the lamellar structure. Therefore, the diverse crystal structural frameworks of layered compounds with exceptional electrochemical capabilities and remarkable electromagnetic phenomena making them pertinent as functional battery materials. Recently, the transition metal compound $Na_2Mn_3O_7$ having a 2D layered crystal structure is reported as a highly efficient battery material with an exceptionally small-voltage hysteresis between the charge and discharge curves (<50 mV), high energy storage capacity, and high structural stability.[12-20] Further studies showed that by an introduction of Mn vacancies in $Na_2Mn_3O_7$ the power capacity can be enhanced up to 220 $mAhg^{-1}$ with a reversible cyclability over 1.5-4.4 V.[14] The compound was also found to work as an efficient cathode material for aqueous Zn ion battery having a power capacity of 245 $mAhg^{-1}$ with a nominal voltage of 1.5 V.[13] Both experimental investigations and DFT calculations showed that during the electrochemical cycling the crystal structure of the compound remained stable with a nominal change in the lattice parameters, i.e., Mn-Mn and Mn-O bond distances.[12, 14, 20] Moreover, during the charging process, the compound showed a Na migration barrier of ~ 0.18 eV, suggesting a good ionic conductivity and a good rate capability facilitated by the lamellar structure of $Na_2Mn_3O_7$.[16] The first-principles molecular dynamics simulations on $Na_2Mn_3O_7$ revealed anisotropic Na ions conductions, where the Na ions conduction occurs within the Na-ion layers with a small fraction of conduction along the perpendicular layers.[14] The spin-charge density analysis revealed that $Mn^{4+}$ ions did not contribute to the conduction process, which was also experimentally verified by scanning transmission electron microscope measurement, where no migration of $Mn^{4+}$ was observed.[14] This implies that the entire conduction in $Na_2Mn_3O_7$ is mediated by Na-ion alone.

Although the technologically important 2D layered compound $Na_2Mn_3O_7$ is extensively investigated for reversible power capacity and charging-discharging properties, a detailed understanding of the ionic conductivity and its microscopic mechanism is yet to be obtained. Knowledge of microscopic conduction process and its mechanism is necessary to design new 2D layered materials with higher efficiency/better performance. For this purpose, the impedance spectroscopy is known to be a powerful technique.[21] In the present work, a comprehensive study on ionic conduction properties including the hopping mechanism of the charge carriers and conductivity relaxation process has been performed by an impedance spectroscopy. The details of the charge transport processes at different lengths and timescales have been obtained from the impedance spectroscopy study. The ionic conduction process is investigated over a temperature range 353-713 K and over a frequency range 1-$10^7$ Hz. The underlying conduction and relaxation mechanisms have been divulged through a combined analysis of dc conductivity, ac conductivity, electric modulus, dielectric constant, and complex polarizability. Crystal structure has been investigated by temperature dependent x-ray diffraction, SEM, and neutron diffraction. The role of crystal structure and stacking faults on the ionic conductivity has been demonstrated by a comprehensive analysis of the x-ray and neutron diffraction patterns on the samples prepared with varying annealing time.

## II. METHODS

The polycrystalline samples of $Na_2Mn_3O_7$ were synthesized by using the standard solid-state reaction method. Stoichiometric mixtures of high purity precursors of sodium nitrate ($NaNO_3$) and manganese carbonate ($MnCO_3$) were thoroughly mixed for ~1 hour using a mortar and a pestle. The stoichiometric mixture was then heated at 600°C for 4 hours in an alumina crucible under an oxygen flow to obtain the single-phase $Na_2Mn_3O_7$ according to the following reaction:

$$2NaNO_3 + 3MnCO_3 + O_2 = Na_2Mn_3O_7 + 3CO_2 + 2NO_2. \quad (1)$$

To investigate the role of sample crystallinity on the ionic conduction, additional samples have been prepared with annealing times of 14 and 28 hours. The densities of the pellets with annealing time of 4, 14 and 28 hours were found to be ~ 88%, 87% and 90%, respectively, of the theoretical density. The conductivity data were corrected for corresponding density of the pellets.

The crystal structural properties of the prepared polycrystalline samples of $Na_2Mn_3O_7$ were investigated over 300-623 K by using a lab source x-ray diffractometer (make: Rigaku, Japan) using a Cu-$K_\alpha$ radiation. The sample was loaded on a platinum plate and mounted on the x-ray diffractometer. The patterns were recorded over the scattering angular range of 10-90 degree with a scan step size of 0.02 degree.

The thermal stability of the compound $Na_2Mn_3O_7$ was studied by thermogravimetric analysis (TGA) and differential scanning calorimetry (DSC) measurements over 303-1024 K. The measurements were performed in the heating cycle at a heating rate of 5 K/min in an Argon (Ar) atmosphere using a thermogravimetric analyzer (make: STA-1600, Linseis).

The surface morphology of the polycrystalline samples of $Na_2Mn_3O_7$ was imaged (magnification up to 25000 times) with a high-resolution field emission scanning electron microscope (FESEM). For these measurements, polycrystalline samples were sprinkled on the top of a carbon-adhesive paper, and thereafter, the loose powders were blown by a hand blower.

Room temperature neutron diffraction measurements were carried out using powder diffractometer-I (λ=1.094Å) at Dhruva research reactor, Mumbai, India. The polycrystalline samples were filled within an 8 mm cylindrical vanadium can and the can was attached to the instrument for the measurements.

The impedance spectroscopy measurements were carried out on a pellet (10 mm diameter and 1.54 mm thickness) sample of $Na_2Mn_3O_7$ using a commercial impedance analyzer PSM1735– NumetriQ (make: Newtons 4th Ltd, UK) over the frequency ($\nu$) range of 1–$10^7$ Hz with an ac electric field of 0.05 V. Silver paste was uniformly coated on the both flat sides of the pellet and dried at 363 K for an hour before the measurements. Prior to the impedance measurements, the pellet was heated at 423 K for an hour to remove moisture. Then, the impedance spectra were recorded in the heating cycle over a temperature range of 353-713 K with a step size of 15 K.

## III. RESULTS

### 1. Crystal Structure (x-ray diffraction):

Figure 1a shows x-ray diffraction (XRD) patterns of $Na_2Mn_3O_7$ at different temperatures over 300-623 K. The diffraction patterns contain a combination of sharp and broad peaks in agreement with the earlier reports.[15, 17, 19] The analysis of the XRD patterns confirms the single-phase formation with the triclinic crystal structure having the space group *P*-1. The crystal structure of $Na_2Mn_3O_7$ is composed of alternating layers of $Mn_3O_7^{2-}$ slabs and Na-ions along the crystallographic [-110] direction (Figure 1b). The $Mn_3O_7^{2-}$ layers are built up with the edge-sharing $MnO_6$ octahedra, and are well separated (~ 5.57 Å) by an intermediate Na-ion layer. In each of such layers, 1/7$^{th}$ of the Mn sites are vacant in a regular manner leading to a maple-leaf lattice geometry of the Mn ions. Within the Na-ion layers, the Na-ions are located at two distinct crystallographic sites having distorted prismatic $Na1O_5$ (situated above the Mn vacant cites) and distorted octahedral $Na2O_6$ environments. This is a rare example of Na coordination environment and called a hybrid prismatic-octahedral (P-O) system.[12, 20] The broad peaks in the XRD patterns appear due to stacking faults of the $Mn_3O_7^{2-}$ layers in the crystal structure of $Na_2Mn_3O_7$.[17-18] The inset of Figure 1a highlights the broad peaks of the measured pattern (points) and compare with the pattern (black line) of the stacking fault free ideal structure of $Na_2Mn_3O_7$ (ICSD_collcode5665). The Bragg peaks (1-10) and (2-20) are sharp whereas, the Bragg peaks (-1-21), (012), (-2-11), (0-23), (-203), (-122), (3-20), (-301), (1-33), (1-41), (2-32), (3-22), and (-411) are broad asymmetric peak shape suggesting the presence of stacking faults (detailed discussion on the stacking faults is given later).

Our temperature-dependent XRD study up to 623 K confirms that the triclinic crystal symmetry with space group *P*-1 remains unchanged. The phase stability of the studied compound $Na_2Mn_3O_7$ is further studied by the TGA and DSC analysis. The temperature-dependent TGA curve (Figure S1) shows (i) a small step (weight loss ~1%) around 373 K due to evaporation of absorbed moisture, and (ii) a big step (weight loss ~ further 6%) at ~750 K due to the decomposition of the sample. The DSC curve (Figure S1) also shows a broad peak around 373 K corresponding to the evaporation of moisture. The TGA/DSC curves, therefore, reveal that the compound $Na_2Mn_3O_7$ is stable up to ~750 K. The study further indicates that the present compound $Na_2Mn_3O_7$ is moisture sensitive and therefore, a special attention/care was taken during the impedance measurements.

The thermal variations of the lattice parameters and unit cell volume derived from the XRD data analysis are shown in Figure 2. With increasing temperature, a nominal increase in the lattice parameters, and the unit cell volume (~ 1.53% over 300-623 K) is observed with a linear volume thermal expansion coefficient of $4.79\times10^{-5}$ $K^{-1}$. The observed thermal expansion coefficient value is comparable to the values reported for other Na/Li-ion battery material compounds.[22-23]

The scanning electron microscope (SEM) images (Figures S2a-c) reveal the micrometre size particles of the polycrystalline samples with an average diameter of ~ 0.4-0.6 µm. The particle size distribution (Figure S2d) is determined by using the ImageJ software by taking into consideration of over 200 particles.[24] The SEM image in Figure S2c further yield that the particles have 2D plate-like shapes due to the large *d*-spacing of the layered structure.

## 2. Electrical Conduction Properties:

Recent studies showed that a parallel RC model serves as a suitable representation for long-range conductivity of a semiconductor/dielectric material.[25] As per the parallel RC model, the electrical conduction, dielectric properties and electrical modulus are inter-connected through the parameters complex impedance ($Z^*$), dielectric constant ($\varepsilon^*$), and electrical modulus ($M^*$) as follows:

$$Z^*(\omega) = \frac{1}{j\omega C_0 \varepsilon_0 \varepsilon^*} = \frac{M^*}{j\omega C_0} \quad (2)$$

where,

$$Z^*(\omega) = Z'(\omega) - jZ''(\omega) = \left(\frac{1}{R} + \mathrm{j}\omega\mathrm{C}\right)^{-1} \quad (3)$$

$$Z'(\omega) = \frac{R}{(1+\omega^2 C^2 R^2)}, \quad (4)$$

$$Z''(\omega) = \frac{\omega C R^2}{(1+\omega^2 C^2 R^2)}, \quad (5)$$

$$\varepsilon*(\omega) = \varepsilon'(\omega) - j\varepsilon''(\omega), \quad (6)$$

and

$$M^*(\omega) = M'(\omega) + jM''(\omega), \quad (7)$$

The parameters $Z'$, $Z''$, $\varepsilon'$, $\varepsilon''$, $M'$, $M''$ are the real and imaginary parts of complex impedance ($Z^*$), dielectric constant ($\varepsilon^*$), and electrical modulus ($M^*$), respectively. $C_0$ is the capacitance of the empty cell, $R$, $C$, $\omega = 2\pi\nu$, $\nu$ and $\varepsilon_0$ are resistance, capacitance, angular frequency, frequency, and free space permittivity, respectively.

In the present study, the impedance spectroscopy measurements were carried out over the frequency (ν) range of 1–$10^7$ Hz and the temperature range of 353-713 K with a temperature step size of 15 K. The recorded impedance data are used to carry out a combined analysis of the ionic conduction and relaxation through the interlinked parameters, viz. dc conductivity, ac conductivity, electric modulus, dielectric constant, and complex polarizability. Below, we discuss the temperature and frequency dependent behaviours of each of these parameters.

### 2.1. DC Conductivity:

Figure 3a shows the frequency dependence of the real ($Z'$) and imaginary ($-Z''$) parts of impedance of the studied compound $Na_2Mn_3O_7$ at various temperatures. At 353 K, the value of $Z'$ is ~ $10^6$ Ohm at 1 Hz and remains almost constant up to ~ $10^3$ Hz suggesting an insulating behaviour. With further increasing frequency above the critical frequency $\nu_c$ ~ $10^3$ Hz, the $Z'$ value reduces sharply. With increasing temperature, $Z'$ value decreases over the whole frequency range, and the critical frequency $\nu_c$ moves towards the higher frequencies. The frequency-dependent imaginary part of impedance, $-Z''$ vs $\nu$ curve at 353 K shows a peak at ~ $10^3$ Hz (Figure 3b). The peak frequency increases with increasing temperature and becomes higher than $10^7$ for $T > 700$ K.

We use the complex impedance spectrum (Nyquist plots or Cole-Cole plot) to estimate the dc-conductivity.[26] The Cole-Cole plots for $Na_2Mn_3O_7$ in Figure 3c show one depressed semicircle as well as an upturn at lower frequencies (higher $Z'$ value region) for all temperatures. Such a depressed semicircle indicates the presence of a relaxation time distribution. The small upturn indicates the presence of a small electrode effect. With increasing temperature, the semicircle decreases sharply and shifts towards higher frequencies.

The dc resistance ($R_{dc}$) values are determined from the intercepts of the semicircles to the $Z'$ axis at the lower frequency side. The dc conductivity $\sigma_{dc}$ values are then calculated from the relation,

$$\sigma_{dc} = \frac{t}{(R_{dc} * A)} \ , \tag{8}$$

where $t$ is the thickness, and $A$ is the cross-sectional area of the pellet. Temperature-dependent $\sigma_{dc}$ (Figure 3d) follows an activation behaviour as

$$\sigma_{dc} = \sigma_0 exp^{-\frac{E_a}{K_B T}} \ , \tag{9}$$

where, $E_a$ is an activation energy, $\sigma_0$ is the exponential pre-factor being the conductivity when the reciprocal temperature approaches zero, $k_B$ is the Boltzman constant and $T$ is the temperature in absolute scale. The linear behaviour of the Arrhenius plot (Figure 3d) reveals a thermally activated conduction process. With increasing temperature, the dc conductivity value increases from ~ $10^{-5}$ $Sm^{-1}$ at 353 K to ~ $10^{-1}$ $Sm^{-1}$ at 713 K. The linear fitting to the curve reveals two different slopes corresponding to activation energies 0.161(7) eV and 0.377(6) eV for the temperature regions 383– 518 K and 518– 713 K, respectively. The derived activation energy $E_a = 0.161$ eV is in good agreement with the value of 0.18 eV predicted by the earlier reported DFT calculation.[16] Here we point out that, the temperature-dependent XRD (Figure 1), TGA and DSC studies (Figure S1) do not show any crystal structural transition up to 700 K. Therefore, the origin of the observed change in activation energy (change in slope in the Arrhenius plot) at ~ 518 K can be ruled out due to the crystal structural phase transition of the compound. The deviation from the linear behaviour of the Arrhenius plot can be explained from several physical properties such as, intrinsic and extrinsic regions of material,[27] transport of charge carriers along the boundaries with ionic conduction, additional scattering, and impurity segmentation, potential fluctuations at the boundary that arises due to an inhomogeneity of the sample,[28] and hydration process.[29] In the present case, the higher activation energy at higher temperature regime indicates that the slope change at ~ 520 K occurs due to the transition from the extrinsic regime to the intrinsic regime with increasing temperature. In this case, the higher activation energy at higher temperature region is due to the combined effect of migration of charge carriers as well as defect formation. At the low temperatures, the activation energy is solely due to migration of charge carriers of existing defects, and, therefore, it has a lower value.[27]

## 2.2. AC Conductivity:

Now we present the ac conductivity properties of the compound. The ac conductivity ($\sigma^*$) is defined as,

$$\sigma^* = \omega \varepsilon_0 \varepsilon^* \tan \delta \ , \tag{10}$$

with the real part,

$$\sigma' = (t/A)\left(Z'/(Z'^2 + Z''^2)\right) , \tag{11}$$

where, $Z'$ and $Z''$ are the real and imaginary parts of the impedance (Figure 3), respectively; the other symbols have the same meaning as before.

The behaviours of the real part of the ac conductivity ($\sigma'$) as a function of frequency and temperature are shown in Figure 4a. At 353 K, the $\sigma'(\nu)$ curve shows a constant value of ~$10^{-5}$ $Sm^{-1}$ up to ~$10^3$ Hz, and then shows a frequency activated behaviour for higher frequencies. With increasing temperature, the frequency-independent value of $\sigma'$ increases monotonically and becomes ~ $10^{-1}$ $Sm^{-1}$ at 713 K. The frequency-independent region also extends to higher frequencies with increasing temperature. According to the relaxation model, at lower frequencies, frequency-independent conductivity is due to a dc conductivity and at higher frequencies, the conduction value increases due to additional frequency activated hopping of charge carriers.

For each temperature, the $\sigma'$ (ν) curves are fitted with the Jonscher's power law,

$$\sigma'(\nu) = \sigma'_{dc} + \alpha\nu^n , \tag{12}$$

where, $\sigma'_{dc}$ is the frequency independent dc conductivity, the temperature dependent pre-factor $\alpha$ represents the polarizability strength of the hopping ions, and the frequency exponent $n$ determines the degree of interactions of charge carriers with surrounding ions, respectively, while performing hopping between two lattice sites.[30] With increasing temperature, the $n$ value initially remains almost constant at ~ 0.5, and then decreases sharply to a value ~ 0.2 at around ~ 518 K (Figure 4b), the temperature where a slope change in the dc conductivity is found (Figure 3d). With further increasing temperature, the $n$ value remains almost constant around 0.2. According to a jump relaxation model,[31-32] the parameter $n$ can be expressed as the ratio between the back-hop rate ($b_r$) and site relaxation rate ($s_r$) as $n = b_r/s_r$. The $b_r$ is the backward motion of a hopping ion to its initial site which is caused by the repulsive Coulomb interaction between mobile ions. The site relaxation is the shift of a site potential minimum to the position of the hopping ion, which is caused by a rearrangement of neighbouring ions. For all temperature ranges, the fitted values of $n$ are found to be much less than the value unity ("1"), revealing that the backward hopping is slower than the site relaxation time, and resulting to a translational motion or conduction of Na-ions. The slower backward motion is due to the smaller coulombic interaction between the ions. The low repulsive coulombic force between mobile ions may result into the low activation energies as found in the dc-conductivity study (Figure 3). The sharp decrease of the $n$ value at ~ 518 K indicates an increase of the hopping range of the charge carriers, where $n$ = 0 corresponds to the dc conductivity. The temperature-dependent $\alpha$ curve initially remains constant up to ~ 518 K (Figure 4c) and then shows a linear temperature dependence for higher temperatures, reflecting the change of hopping range at ~ 518 K.

Temperature-dependent $\sigma'$ curves for selective frequencies (Figure 5a) show enhancement of the conductivity ($\sigma'$) values implying the semiconducting behavior of the compound. The dc conductivity ($\sigma'_{dc}$) values (@ 1 Hz) are in well agreement with the $\sigma_{dc}$ values derived from the Cole-Cole plot of impedances (Figure 3c). The temperature dependence of $\sigma'$ for all frequencies follows the Arrhenius behaviour;

$$\sigma' = \sigma'_0 exp^{-\frac{E'_a}{K_B T}} . \tag{13}$$

Figure 5b shows the Arrhenius behaviour of $\sigma'$ for a representative frequency of $10^5$ Hz. The activation energies ($E'_a$) for selected frequencies, determined from the $\sigma'$ vs $T$ curves, are given in Table 1. It is found that the activation energies ($E'_a$) for both the temperature ranges (above and below 518 K) initially remain constant up to a critical frequency $\nu_C$ and then decrease with further increasing frequency value above the critical frequency $\nu_C$ (Figure 5c), revealing a frequency activated enhancement of the ionic conductivity.[33] We have shown that despite a significant enhancement of the $\sigma'_{dc}$ value (~ $10^4$ times) (Figure 4a) with increasing temperature from 353 K to 713 K, the $\sigma(\nu)$ curves at different measurement temperatures can be scaled by ($\sigma'(\nu)/\sigma'_{dc}$) and ($\nu/\nu_C$) (Figure 5d). This indicates that the frequency activated conduction is mainly controlled by the critical frequency ($\nu_C$) whose value increases with the increasing temperature.

### 2.3. Electrical Modulus:

The conduction properties of $Na_2Mn_3O_7$ are further investigated by the electrical modulus formalism, an important tool to extract useful information about the conductivity relaxation phenomenon of the material under investigation. The electrical modulus physically attributes to the relaxation of the electric field in a compound considering a constant electric displacement. The impedance and modulus formalism are used together as a crucial tool to study the electrical response in ionic conductors where long-range movement of charge carriers dominates.[34] The advantage of electrical modulus spectrum is that it does not contain the electrode effects (unlike impedance). The conductivity relaxation process is manifested by the appearance of peaks in the frequency-dependent imaginary part of the electrical modulus curve. The values of real ($M'$) and imaginary ($M''$) parts of the electrical modulus ($M^*$) are derived from Eqs. 2 and 7 as,

$$M' = -Z'' \times \left(\frac{\omega A \varepsilon_0}{t}\right) , \tag{14}$$

and

$$M'' = Z' \times \left(\frac{\omega A \varepsilon_0}{t}\right) , \tag{15}$$

Figure 6a represents the comparative plots between frequency-dependent imaginary part of the impedance ($Z''$), real ($M'$) and imaginary ($M''$) parts of the electrical modulus measured at 353 K. The peaks in the $Z''(\nu)$ and $M''(\nu)$ curves at ~ $10^3$ Hz almost coinciding with each other revealing the conductivity relaxation process. The weak cusp in the $M''(\nu)$ curve at a higher frequency of ~ $10^5$ Hz may arise due to a dipole relaxation. Both $M'(\nu)$ and $M''(\nu)$ approach to zero for low frequencies indicating the suppression of electrode effect in the modulus spectra. With increasing frequency, there is a sharp increase in the $M'$ value at a critical frequency and thereafter a saturation at high frequencies (Figure 6b). With increasing temperature, the overall $M'$ value decreases, and the critical frequency shifts towards the higher frequencies. With increasing temperature, the peak in the $M''$ curves shifts towards high frequencies (Figure 6c), suggesting that the conductivity relaxation process is thermally activated and it is dominated by hopping of the charge carriers.[35] The shift of the peak towards higher frequency means a reduction in the relaxation time, hence, a faster ionic motion.

For an ionic conduction process, the modulus curves can be represented by the Kohlrausch–Williams–Watts (KWW) decay function. The frequency-dependent imaginary part of the modulus ($M''$) curve can be expressed by the KWW function as:

$$M'' = \frac{M''_{max}}{\left[(1-\beta)+\left(\frac{\beta}{1+\beta}\right)\left(\beta\left(\frac{\nu_{max}}{\nu}\right)+\left(\frac{\nu}{\nu_{max}}\right)^{\beta}\right)\right]} , \tag{16}$$

where, $M''_{max}$ is the maximum value of $M''$ at the frequency $\nu_{max}$, and $\beta$ is the stretched exponential parameter. The value of $\beta$ lies between $0 < \beta < 1$ and represents the deviation from the linear exponential i.e., the deviation from the Debye type relaxation (where $\beta = 1$ corresponds to Debye relaxation).[36] All $M''(\nu)$ curves measured over the entire temperature range follow the KWW decay function as in Eq. 16 (Figure 6c). The temperature dependence of the fitted values of the $\beta$ is shown as the inset in Figure 6c. The value of $\beta$ at 353 K is found to be 0.72 which remains almost constant upto 518 K and then increases with increasing temperature and approaches to unity at ~ 700 K. With the increasing temperature, the increase of $\beta$ values suggests that the relaxation process shifts towards the Debye-type, and the corresponding relaxation time distributions become narrower.

The Cole-Cole plots in the electric modulus plane ($M''$ vs $M'$) (Figure 6d) reveal a single depressed semicircle for the curves over the entire temperature range as found for the Cole-Cole plots of the impedances ($-Z''$ vs $Z'$) (Figure 3c). Besides, the imaginary part of electrical modulus ($M''$) curves show a universal scaling behaviour ($M''/ M''_{max}$ vs $\nu/\nu_{max}$) over the whole temperature range (Figure 6e). The normalized curves (master curves) merge into each other revealing the universal feature of the relaxation dynamics. This implies that in $Na_2Mn_3O_7$ the conductivity relaxation process has the same mechanism over the entire measurement temperature range of 383-713 K.[37]

The conductivity relaxation time can be calculated from the peak frequencies in the $-Z''(\nu)$ or $M''(\nu)$ curves by using the relation,

$$\omega_m \tau_m = 2\pi \nu_m \tau_m = 1 \ , \tag{17}$$

where, $\nu_m$ is the frequency at the relaxation point where $-Z''$ and $M''$ are maximum. The relaxation time ($\tau_m$) is found to be $2.68\times10^{-4}$ sec at 353 K, and the $\tau_m$ value decreases to $2.68\times10^{-7}$ sec with increasing temperature to 713 K (Figure 6f), suggesting a faster conductivity relaxation at higher temperatures. The temperature dependence relaxation time, $\tau_m$ follows the Arrhenius behaviour,

$$\tau_m = \tau_0 exp^{\frac{Ea}{K_B T}} \ , \tag{18}$$

where, $\tau_0$ is the characteristic relaxation time. The values of $\tau_0$ are determined to be $3.6\times10^{-9}$ and $2.38\times10^{-13}$ for the temperature regions over 383-518 K and 518-713 K, respectively. The activation energy is found to be 0.162(4) eV and 0.347(4) eV for temperature ranges 383-518 K and 518-713 K respectively (Figure 6f). Both the dc-conductivity (Figure 3d) and conductivity relaxation time (Figure 6f) show the Arrhenius behavior with close values of the activation energies, reflecting the fact that dc conductivity and conductivity relaxation time are conjugate to each other and related by the equation,

$$\tau_m = \frac{\varepsilon_0 \varepsilon_s}{\sigma_{dc}} \ , \tag{19}$$

where, $\varepsilon_0$ and $\varepsilon_s$ are free space permittivity and static dielectric constant, respectively. This also implies that the charge carrier has to overcome the same energy barrier for the conduction as well as relaxation processes. The interesting fact is that the relaxation time decreases by three orders of magnitude with an increase of the temperature from 383 K to 713 K which is a typical signature of the thermally activated conduction process.

**2.4. Dielectric Constant:**

The typical feature of the dielectric constant is useful to understand the inherent local dipole relaxation properties. The real ($\varepsilon'$) and imaginary ($\varepsilon''$) parts of the dielectric constant ($\varepsilon^*$) are derived from Eqs. 2 and 6 as,

$$\varepsilon'(\omega) = \frac{t}{\omega A \varepsilon_0}\left(\frac{Z''}{Z'^2+Z''^2}\right) , \tag{20}$$

and

$$\varepsilon''(\omega) = \frac{t}{\omega A \varepsilon_0}\left(\frac{Z'}{Z'^2+Z''^2}\right) , \tag{21}$$

The real part of dielectric constant ($\varepsilon'$), associated with electric polarization, shows a high value ~$10^4$ at 1 Hz and 353 K (Figure 7a). With increasing frequency, the $\varepsilon'$ value decreases exponentially and approaches to a value ~ $10^2$ at a critical frequency ~$10^3$ Hz and then remains constant for higher frequencies. The increase of the $\varepsilon'$ values at a lower frequency range can be associated with the electrode polarization effect. With increasing temperature, the value of $\varepsilon'$ increases significantly up to ~ $10^6$ (@ 1 Hz) at 713 K (Figure 7a). The value of critical frequency also increases with increasing temperature and approaches ~ $10^6$ Hz at 713 K. All the $\varepsilon'(\nu)$ curves measured over 353-713 K merge to each other at higher frequencies above ~ $10^6$ Hz. The imaginary part of dielectric constant ($\varepsilon''$) is associated with the electrical energy loss. The log-log plots of the frequency dependence of the imaginary part of the dielectric constant $\varepsilon''(\nu)$ at various temperatures show linear behaviour (Figure 7b). The small cusp at ~ $10^5$ over low temperatures may be due to the dipole relaxation losses. The frequency dependence of the dielectric constant $\varepsilon''$ has been found to follow the universal power law,

$$\varepsilon'' = B\omega^m , \tag{22}$$

where, $B$ is a proportional constant and $m$ is a temperature-dependent exponent. With increasing temperature, the $m$ value decreases and approaches to the value "-1". The decrease in $m$ value with temperature suggests a predominantly correlated barrier hopping (CBH) process in the present compound.[38-39]

Based on the above experimental observations from impedance, electrical modulus and dielectric constant, we clarify below that the ionic conduction is the dominating process in the studied compound $Na_2Mn_3O_7$. For a pure conduction process, a relaxation peak occurs in the frequency-dependent $M''$ spectrum, whereas, no peak appears in the frequency-dependent $\varepsilon''$ spectrum. On the other hand, for a localized dipole relaxation process, relaxation peaks are observed in both the frequency-dependent $M''$ and $\varepsilon''$ spectra.[40] For the studied compound $Na_2Mn_3O_7$, a relaxation peak is found in the frequency-dependent $M''$ curve at the same position of the peak in the $Z''(\nu)$ curve (Figure 6a), whereas, no corresponding peak is found in the frequency-dependent $\varepsilon''$ curve (Figure 7b) revealing a dominated ionic conduction process. Moreover, the narrow peak width of the frequency-dependent $M''$ curves suggests an ionic conduction process.[41] Therefore, it may be concluded that the Na-ionic conduction is the dominating process in the present compound $Na_2Mn_3O_7$.

## 2.5. Complex Polarizability:

The dielectric response has further been analyzed by a new method, proposed by Scaife,[42] known as complex polarizability ($\alpha^*$). The parameter $\alpha^*$ is especially advantageous to investigate the intrinsic dielectric properties of a material as it provides a proper weightage to all polarization mechanisms. Therefore, multiple coexisting relaxation processes in a material can be separated out more efficiently through this formalism. The complex polarizability parameter ($\alpha^*$) is defined by:

$$\alpha^* = \alpha' - i\alpha'' = \frac{(\varepsilon^*-1)}{(\varepsilon^*+2)} , \quad (23)$$

where, $\alpha'$, $\alpha''$, are real and imaginary parts of the complex polarizability ($\alpha^*$), respectively, and $\varepsilon^*$ is the complex dielectric constant. Using the above relation, the imaginary part of the complex polarizability parameter $\alpha''$ can be calculated as follows:

$$\alpha'' = \frac{3\varepsilon''}{(\varepsilon'+2)^2+\varepsilon''^2} , \quad (24)$$

where, $\varepsilon'$ and $\varepsilon''$ are real and imaginary parts of the dielectric constant ($\varepsilon^*$), respectively. Figure 8a shows the frequency-dependent $\alpha''$ curves at different temperatures. Each of the $\alpha''(\nu)$ curves show a well-defined relaxation peak with a maximum at $\nu_\alpha$. With increasing temperature, the relaxation peak shifts towards the higher frequencies and follows an Arrhenius behaviour;

$$\nu_\alpha = \nu_{\alpha 0} exp^{\left(\frac{-E_a}{k_B T}\right)} , \quad (25)$$

where, $E_a$ is the activation energy. The Arrhenius plots of the peak frequencies $\nu_\alpha$ and $\nu_m$ of the imaginary parts of polarizability ($\alpha''$) and modulus ($M''$) overlap with each other for the entire temperature range (Figure 8b), suggesting a thermally activated conduction process. The activation energies are found to be 0.162(5) eV and 0.347(6) eV for the temperature ranges 383-518 K and 518-713 K, respectively. The activation energies are the same as the values obtained from the dc-conductivity and the electric modulus studies. This result, hence, concludes that the thermally activated Na ion conduction in the present compound $Na_2Mn_3O_7$ occurs through the hopping process.

### 2.6. Dielectric Loss Factor (*tanδ*):

The loss of electrical energy due to the hopping of charge carriers and orientation of electric dipoles is represented by the dielectric loss factor. The dielectric loss factor is defined as,

$$tan\delta = \frac{\varepsilon''}{\varepsilon'} = \frac{M''}{M'} = \frac{Z'}{Z''} \quad (26)$$

Figure 9a shows that the value of loss factor *tanδ* is relatively low around 10-40 at lower frequency range. The value of *tanδ* decreases rapidly with increasing frequency and becomes zero for higher frequencies. The increase of the hopping of charge carriers with increasing frequency results into the reduction of the loss factor at a higher frequency range. The observed low values of *tan*δ in $Na_2Mn_3O_7$ correspond to a low dissipative energy. Such low energy loss materials are desired for battery applications. A small hump appears in the *tanδ* ($\nu$) curves at $\nu_D$ ~10 Hz at 353 K. The scaling behavior of the loss factor ($tan\delta/tan\delta_{max}$ vs $\nu/\nu_{max}$) is shown in (Figure 9b), where $tan\delta_{max}$ represents the peak value of *tanδ* at the frequency $\nu_{max}$. The normalized curves for all the temperatures merge into a master curve, as like the modulus curves (Figure 6e), establishing the temperature-independent universal feature of the relaxation dynamic processes. The coincidence of the peaks of loss factor also yields that the conductions of Na ions in $Na_2Mn_3O_7$ take place through the hopping mechanism.[38]

In summary, the above analyses of the impedance data (sections: 2.1-2.6) in the form of dc-conductivity, ac-conductivity, electrical modulus, dielectric constant and complex polarizability, therefore, reveal a long-range Na-ionic conductivity with a negligible contribution from the dipole relaxation. Our analyses also reveal that the thermally activated Na-ionic conductions in $Na_2Mn_3O_7$ occur through the correlated barrier hopping process.

### 3. Crystal Structure (Neutron diffraction):

Neutron diffraction patterns at room temperature for the 4, 14, and 28 hours annealed samples are shown in Figure 10. The measured patterns contain both sharp and broad Bragg peaks. The profiles of the broad peaks are asymmetric. The diffraction pattern for an ideal crystal structure of $Na_2Mn_3O_7$ without stacking faults (ICSD_collcode5665) was also simulated and shown in Fig. 10a. With the increasing annealing time substantial changes are evident for the (012), (102) and (-203) Bragg peaks (marked with asterisks). The changes of peak profile parameters (i.e., peak intensity, peak width, peak asymmetry, and peaks shift) with the annealing time have been quantitatively estimated by fitting the observed neutron diffraction patterns over the 2θ range ~ 23.5-37 degree by seven peaks [as expected from the crystal symmetry of $Na_2Mn_3O_7$ (Figure 10a)] each having the following asymmetric PearsonIV peak function,

$$y = k[1 + (x - \lambda/_{\alpha})^2]^{-m} exp^{-v tan^{-1}(x-\lambda/_{\alpha})} \quad (27)$$

Where, $k = \frac{2^{2m-2}|\Gamma(m+iv/_2)|^2}{\pi\alpha\Gamma(2m-1)}$, α = peak width, λ = peak centre, and v = asymmetry parameter.

The variations of the integrated intensity, peak width, asymmetric parameter and peak shift for the (012), (102), and (-203) Bragg peaks with annealing time are shown in Figures 10e-h. The variation of such peak profile parameters is also compared with the same for the unaffected Bragg peak (3-30) (Figures 10e-h). With the increasing annealing time, an increase in the peak intensity whereas a decrease in the peak width, peak asymmetry, and peaks shift are found for (012), (102), and (-203) Bragg peaks. The derived results are quite significant to establish the role of crystal structure on the ionic conductivity as presented in next section.

## IV. DISCUSSIONS

Now we shed light on the conduction mechanism and relaxation process in the studied 2D layered compound $Na_2Mn_3O_7$. For the ac-conduction mechanism, two main types of processes have been proposed viz., (i) quantum mechanical tunnelling (QMT) [43-44] and (ii) classical hopping over a barrier (HOB) models.[45-46] In the QMT model, the conduction of ions occurs through quantum mechanical tunnelling between localized (defected) states. As per this model, the frequency exponent $n$ is independent of temperature with a constant value of ~ 0.81.[47] However, for the studied compound $Na_2Mn_3O_7$, our experimental results yield that the $n$ values are smaller (< 0.55) and decrease with increasing the temperature (Figure 4), hence, rule out the applicability of this model. On the other hand, within the classical HOB model, the value of $n$ is predicted to be equals to unity. The much smaller experimentally derived values of the $n$ for $Na_2Mn_3O_7$, therefore, indicate the correlated barrier hopping (CBH) model. In this model, the motion of charge carriers occurs through hopping over the Coulomb barrier which results in a correlation between the relaxation variable $W$, of the coulomb barrier and the intersite separation $R$. The CBH model predicts that the frequency exponent $n$ is temperature dependent and the $n$ decreases with the increasing temperature. The value of $n$ becomes unity as the temperature tends to unity. Further, our temperature and frequency-dependent imaginary part of dielectric constant ($\varepsilon''$) data (Figure 7) suggest the correlated barrier hopping process.

Now we discuss the ion relaxation mechanism of $Na_2Mn_3O_7$ which is found to be remarkedly non-Debye type (Figure 3 and 6), especially at lower temperatures. For such a behaviour, in general there are two possible origins based on (i) parallel or (ii) series processes.[48] In the case of a parallel relaxation, each microscopic process may follow a Debye behaviour, but a wider distribution of relaxation times exists as a consequence of the disordered state. On the other hand, for the case of series processes, individual relaxation processes are coupled and each relaxation event is constrained by the others. During each elementary step of translational ionic motion, a mismatch between the ion and its neighbourhood is generally created. The mismatch is a measure of the distance between the actual position of the ion just after the hopping and the final relaxed position with respect to the momentary arrangement of its mobile neighbours. There are two possible ways of mismatch relaxation, viz., (i) the single-particle route, and (ii) the many-particle route. In the case of a single-particle route, the ion itself performs a correlated backward movement, while in the second case, a rearrangement of the neighbouring ions occurs. These single-particle and multiparticle motions are defined by a time-dependent correlation factor $W(t)$ and a normalized mismatch relaxation function $g(t)$, respectively.[21] The function $W(t)$ is a measure of the time-dependent average normalized displacement of an ion caused by elementary steps. At $t$=0, $W(t)$ is normalized to one, and then it decays with time to its final value $W(\infty)$ which is the fraction of successful hops and called the correlation factor. Similarly, the mismatch relaxation function $g(t)$ is also normalized to one when $t$=0. At finite time, the function $g(t)$ describes the relaxational decay of the mismatch due to the movements of neighbouring ions, and becomes zero in the limit of long times $t\rightarrow\infty$. It should be mentioned that the rates of decay of $W(t)$ and $g(t)$ are not independent, but synchronously coupled to each other.[21, 31] From our ac conductivity results (Figure 4) we estimate the value of $W(\infty)$ ~ 0.45 at 353 K which enhances to the value of $W(\infty)$ ~0.8 with the increasing temperatures to 713 K.

We discuss below the role of crystal structure on the conduction and battery properties of $Na_2Mn_3O_7$. The advantage of the intrinsic regular Mn vacancies (1/7$^{th}$ of the Mn sites) in the Mn-O slabs of the crystal structure of the compound $Na_2Mn_3O_7$ is that the high structural flexibility, i.e., the rigidity of the host structure and its stability during an electrochemical cycling. The nonbonding 2*p* orbitals of oxygen ions that are neighbouring the Mn vacancies contribute to the oxygen-redox capacity without making any change to the Mn−O bond, leading to the highly reversible oxygen-redox capacity of $Na_2Mn_3O_7$ with the exceptionally small-voltage hysteresis between the charge and discharge processes.[12] The de-intercalation and intercalation of sodium ions only affect the dimension of the Mn-O plane in the crystal structure without any noticeable change of the interlayer spacing and the overall oxygen ions stacking sequences. This prevents the irreversible P-type to O-type phase transition in $Na_2Mn_3O_7$ in contrast to the other P-type layered sodium oxide cathode materials during an electrochemical cycling. Another important crystal structural feature is that the large interlayer spacing of ~ 5.57 Å is occupied by $Na^+$ ions alone and facilitates the Na-ion conduction. Unlike the layered $Na_xMnO_2$ compound with O3-type crystal structure, the vacant Mn-sites in the present compound $Na_2Mn_3O_7$ provide some ionic conduction perpendicular to the layers. On the other hand, the site disorder or the mixed occupations of the Na and Mn sites are minimally expected for the studied compound. The large difference in the ionic radius of $Na^+$ (~1.0 Å) and $Mn^{4+}$ (0.53 Å) effectively prevents the Mn-ion migration to the Na sites. The DFT study on $Na_2Mn_3O_7$ revealed that a large energy is required for the migration of Mn ions into the Na layers due to the high structural stability, and such migration will lead to phase transitions from the layered structure to a spinel crystal structure.[14] As a result, such Mn ions migrations are minimally expected in $Na_2Mn_3O_7$ under normal conditions. Therefore, the observed broadening of Bragg peaks in the XRD patterns (Figure 1) cannot be correlated to the site disorders. Rather, stacking faults of the layers are highly expected for the studied compound $Na_2Mn_3O_7$ as often found for the related layered battery materials.[49-50] This is explored by investigating samples that are prepared with varying annealing time (4, 14, and 28 h). It is expected that with the increasing annealing time the stacking faults can be reduced. Figure 10a shows the neutron diffraction pattern for the 4 hours annealed sample along the calculated pattern for the ideal crystal structure without stacking faults (ICSD_collcode5665). As like xrd pattern (Figure 1), the neutron diffraction pattern contains both sharp and broad Bragg peaks. Especially, the Bragg peaks which are indexed with (h-k0), i.e., (1-10) (2-20), and (3-30) are sharp, whereas, the Bragg peaks (012), (102), (-203), (0-31) and (2-12) are broad. The profiles of the broad peaks are asymmetric [Figures 1 and 10a-d]. Similar types of diffraction patterns with a combination of sharp and broad peaks are reported for several layered compounds and such features in the diffraction patterns are explained by stacking faults of the layered crystal structures.[51-54] The stacking faults are planar defects that can be described as stacking of different layer types. Besides, (2-12), (102) and (012) Bragg peak positions are shifted as compared to the pattern for the stacking fault free structure. The above results indicate that the stacking faults cause not only line broadening (anisotropic), but also line shifts.

For the present compound $Na_2Mn_3O_7$, within the unit cell, there are two manganese-oxygen $Mn_3O_7^{2-}$ layers which are approximately parallel to the (1-10) planes (Figure 1). The layers are shifted by (a/2,b/2,0) with respected to each other (Figure 1). The crystal structure is similar to the T2 type crystal structure reported for the layered battery materials $AMO_2$ (where, A=Li/Na, M = Ni/Mn/Co).[51] The Bragg peaks (*h*-*k*0), such as (1-10) (2-20), and (3-30), originate from the $Mn_3O_7^{2-}$ manganese-oxygen layers, whose position and width are not affected by stacking faults suggest that the periodicity of the layers remains intact along the perpendicular direction. We stress upon the fact that these Bragg peaks appear at the same positions in the diffraction pattern as expected for a stacking free ideal crystal structure of $Na_2Mn_3O_7$ (ICSD_collcode5665) revealing that the interlayer spacing in the present compounds is not affected by the stacking faults. On the other hand, the broadening of the Bragg peaks (-1-13), (012), (102), and (-203) [i.e., the Bragg peaks originating from the planes non-parallel to the manganese-oxygen layers] clearly yields the presence of stacking faults of the manganese-oxygen $Mn_3O_7^{2-}$ layers. The stacking faults are created when the regular stacking sequence of the manganese-oxygen $Mn_3O_7^{2-}$ layer is lost due to an inplane sliding of a particular layer.

The neutron diffraction patterns for the 4, 14, and 28 hours annealed samples are shown in Figures 10b, c, and d, respectively. As evident from the neutron diffraction patterns the effect of stacking faults are best visible on the Bragg peaks that are situated over the 2θ range ~ 24-33 deg. With the increasing annealing time substantial changes are evident for the (012), (102) and (-203) Bragg peaks (marked with asterisks). The changes of peak profile parameters (i.e., peak intensity, peak width, peak asymmetry, and peaks shift) with the annealing time have been quantitatively estimated by fitting the observed neutron diffraction patterns by the asymmetric PearsonIV peak function. The variations of the integrated intensity, peak width, asymmetric parameter and peak shift for the (012), (102), and (-203) Bragg peaks with annealing time are shown in Figures 10e-h. The variation of such peak profile parameters is also compared with the same for the unaffected Bragg peak (3-30) (Figures 10e-h). With the increasing annealing time, an increase in the peak intensity whereas a decrease in the peak width, peak asymmetry, and peaks shift are found for (012), (102), and (-203) Bragg peaks. The above results reveal a reduction of the stacking faults with the increasing annealing time.

In order to investigate the effect of stacking faults on the ionic conductivity, we have performed impedance study on the three samples prepared with 4, 14 and 28 hours annealing times. Figure 11 shows comparative plot of density normalised ionic conductivity of the three samples prepared with 4, 14 and 28 hours annealing times. Our impedance results (Figure 11) show that the ionic conductivity is increasing with increasing annealing time i.e., with the reduction of stacking faults. Therefore, our study established a correlation between the stacking faults and the ionic conductivity where the conductivity increases with the reducing stacking fault of the crystal structure. Similar results i.e., the enhancement of the conductivity with the reduction of the stacking faults are reported for the other layered compound

Li2RuO3.[54] Therefore, the present work would be useful as a reference to tune ionic conductivity of a material by varying the synthesis condition.

We now compare the temperature-dependent conductivity curve of the studied 2D layered compound $Na_2Mn_3O_7$ with the same of the recently studied related Na based layered transition metal oxide materials in Figure 12. The conductivity of $Na_2Mn_3O_7$ is found to be highest among the compounds having the same formula $Na_2M_3O_7$ and similar type layered crystal structure, viz. $Na_2Ti_3O_7$ and $Na_2W_2O_7$. In general, the studied compound $Na_2Mn_3O_7$ shows modest conductivity amongst the layered Na-ion battery materials. Moreover, the present compound shows higher energy storage capacity (>220 $mAhg^{-1}$) as compared to the other reported layered oxide compounds, such as $LiCoO_2$, $Li[Ni_{1-x-y}Mn_xCo_y]O_2$, $LiFePO_4$, $LiMn_2O_4$, and $NaMO_2$ ($M$=Ti, V, Cr, Mn, Fe, Co, Ni, and a mixture of 2 or 3 elements) (having the capacity <200 $mAhg^{-1}$) which are presently used in commercial batteries.[9, 11] The present study further reveals that a low dissipative energy is involved in $Na_2Mn_3O_7$ due to small polarization.[55] The interesting fact is that $Na_2Mn_3O_7$ possesses a very small hysteresis voltage (<40 mV) between the charging and discharging which leads to a high energy efficiency.[12]

## V. CONCLUSIONS:

In summary, detailed investigations of crystal structure and Na-ionic conduction properties of the highly efficient 2D layered battery material $Na_2Mn_3O_7$ have been carried out as a function of temperature. The compound $Na_2Mn_3O_7$ has a layered crystal structure where Na-ion layers and magnetic $Mn_3O_7^{2-}$ layers are arranged alternatively. The detailed x-ray and neutron diffraction study and their analysis reveal the presence of stacking faults. Our study further shows that the stacking faults can be reduced by enhancing the annealing time during sample synthesis. The thermal variation of the unit cell volume has a nominal change (~1.53%) with a thermal expansion coefficient of $4.79\times10^{-5}$ $K^{-1}$. The crystal structural symmetry remains invariant over the temperature range of 353-713 K. A comprehensive investigation of the conduction properties by an impedance spectroscopy reveals that the dc conductivity value increases nearly ~$10^4$ times with the increasing temperature from 353 K to 713 K. Further, the ionic conductivity is found to get frequency activated above the critical frequency $\nu_C$ ($\nu_C$ =$10^4$ Hz at 353 K). Our scaling analysis reveals that the frequency-activated conductivity is mainly controlled by the critical frequency ($\nu_C$) that increases with the increasing temperature. The temperature-dependent conductivity curves for all frequencies indicate two temperature regions with the characteristic activation energies of 0.161 and 0.377 eV (for dc conduction) over the temperature ranges 383– 518 K and 518– 713 K, respectively. Two different hopping ranges of the charge carriers are found over the two temperatures regions of 383–518 K and 518– 713 K. With the increasing frequency above the $\nu_C$, the activation energies for both the regions decrease. The electrical modulus formalism provides a detailed understanding of the conduction relaxations and reveals a thermally activated process. The modulus and the dielectric studies further reveal that the ionic conduction is the most dominating process with low energy loss. The contribution of dipolar relaxation over the studied frequency range is insignificant. The conductivity relaxation time is found to be of the order of $10^{-4}$-$10^{-7}$ seconds suggesting a faster ionic motion. The present comprehensive study through combined analyses of dc- and ac-conductivity, electric modulus, dielectric constant, and complex polarizability facilitates the understanding of the microscopic mechanism of the Na-ions conduction and relaxation processes in the technologically important layered battery material $Na_2Mn_3O_7$. We have further shown that the conductivity of $Na_2Mn_3O_7$ can be enhanced by reduction of the stacking faults in the crystal structure. Our results provide an insight into the fundamental conduction and the relaxation mechanisms in the family of the 2D layer Na-ion based transition metal oxide, paving way for the discovery of novel materials for functional battery applications.

## ACKNOWLEDGEMENT

The authors acknowledge the help provided by V. B. Jayakrishnan, J. Bahadur, and P. Bhatt for the high-temperature x-ray diffraction, SEM, and TGA/DSC measurements, respectively. B.S. thanks the Department of Science and Technology (DST), Government of India, for providing the INSPIRE fellowship (INSPIRE Grant. No. IF180105).

## REFERENCES:

1. Yabuuchi, N.; Kubota, K.; Dahbi, M.; Komaba, S., Research Development on Sodium-Ion Batteries. *Chemical Reviews* **2014**, *114*, 11636-11682.
2. Yang, J.; Zhou, X.; Wu, D.; Zhao, X.; Zhou, Z., S-Doped N-Rich Carbon Nanosheets with Expanded Interlayer Distance as Anode Materials for Sodium-Ion Batteries. *Advanced Materials* **2017**, *29*, 1604108.
3. Ellis, B. L.; Lee, K. T.; Nazar, L. F., Positive Electrode Materials for Li-Ion and Li-Batteries. *Chemistry of Materials* **2010**, *22*, 691-714.
4. Byles, B. W.; Palapati, N. K. R.; Subramanian, A.; Pomerantseva, E., The Role of Electronic and Ionic Conductivities in the Rate Performance of Tunnel Structured Manganese Oxides in Li-Ion Batteries. *APL Materials* **2016**, *4*, 046108.
5. Reddy, M. V.; Subba Rao, G. V.; Chowdari, B. V. R., Metal Oxides and Oxysalts as Anode Materials for Li Ion Batteries. *Chemical Reviews* **2013**, *113*, 5364-5457.
6. Vijaya Babu, K.; Seeta Devi, L.; Veeraiah, V.; Anand, K., Structural and Dielectric Studies of $LiNiPO_4$ and $LiNi_{0.5}Co_{0.5}PO_4$ Cathode Materials for Lithium-Ion Batteries. *Journal of Asian Ceramic Societies* **2016**, *4*, 269-276.
7. Delmas, C., Sodium and Sodium-Ion Batteries: 50 Years of Research. *Advanced Energy Materials* **2018**, *8*, 1703137.
8. Kubota, K.; Dahbi, M.; Hosaka, T.; Kumakura, S.; Komaba, S., Towards K-Ion and Na-Ion Batteries as "Beyond Li-Ion". *The Chemical Record* **2018**, *18*, 459-479.
9. Nayak, P. K.; Yang, L.; Brehm, W.; Adelhelm, P., From Lithium-Ion to Sodium-Ion Batteries: Advantages, Challenges, and Surprises. *Angewandte Chemie International Edition* **2018**, *57*, 102-120.
10. Ma, C.; Alvarado, J.; Xu, J.; Clément, R. J.; Kodur, M.; Tong, W.; Grey, C. P.; Meng, Y. S., Exploring Oxygen Activity in the High Energy P2-Type $Na_{0.78}Ni_{0.23}Mn_{0.69}O_2$ Cathode Material for Na-Ion Batteries. *Journal of the American Chemical Society* **2017**, *139*, 4835-4845.
11. Han, M. H.; Gonzalo, E.; Singh, G.; Rojo, T., A Comprehensive Review of Sodium Layered Oxides: Powerful Cathodes for Na-Ion Batteries. *Energy & Environmental Science* **2015**, *8*, 81-102.
12. Song, B.; Tang, M.; Hu, E.; Borkiewicz, O. J.; Wiaderek, K. M.; Zhang, Y.; Phillip, N. D.; Liu, X.; Shadike, Z.; Li, C.; Song, L.; Hu, Y. Y.; Chi, M.; Veith, G. M.; Yang, X. Q.; Liu, J.; Nanda,

J.; Page, K.; Huq, A.,Understanding the Low-Voltage Hysteresis of Anionic Redox in $Na_2Mn_3O_7$. *Chemistry of Materials* **2019**, *31*, 3756-3765.
13. Sada, K.; Barpanda, P., Layered Sodium Manganese Oxide $Na_2Mn_3O_7$ as an Insertion Host for Aqueous Zinc-Ion Batteries. *MRS Advances* **2019**, *4*, 2651-2657.
14. Li, Y.; Wang, X.; Gao, Y.; Zhang, Q.; Tan, G.; Kong, Q.; Bak, S.; Lu, G.; Yang, X. Q., Gu, L., Native Vacancy Enhanced Oxygen Redox Reversibility and Structural Robustness. *Advanced Energy Materials* **2019**, *9*, 1803087.
15. Sada, K.; Senthilkumar, B.; Barpanda, P., Layered $Na_2Mn_3O_7$ as a 3.1 V Insertion Material for Li-Ion Batteries. *ACS Applied Energy Materials* **2018**, *1*, 6719-6724.
16. Zhang, Z.; Wu, D.; Zhang, X.; Zhao, X.; Zhang, H.; Ding, F.; Xie, Z.; Zhou, Z., First-Principles Computational Studies on Layered $Na_2Mn_3O_7$ as a High-Rate Cathode Material for Sodium Ion Batteries. *Journal of Materials Chemistry A* **2017**, *5*, 12752-12756.
17. Adamczyk, E.; Pralong, V., $Na_2Mn_3O_7$: A Suitable Electrode Material for Na-Ion Batteries? *Chemistry of Materials* **2017**, *29*, 4645-4648.
18. Raekelboom, E. A.; Hector, A. L.; Owen, J.; Vitins, G.; Weller, M. T., Syntheses, Structures, and Preliminary Electrochemistry of the Layered Lithium and Sodium Manganese(Iv) Oxides, $A_2Mn_3O_7$. *Chemistry of Materials* **2001**, *13*, 4618-4623.
19. Chang, F. M.; Jansen, M., Darstellung Und Kristallstruktur Von $Na_2Mn_3O_7$. *Zeitschrift für anorganische und allgemeine Chemie* **1985**, *531*, 177-182.
20. Mortemard de Boisse, B.; Nishimura, S.-i.; Watanabe, E.; Lander, L.; Tsuchimoto, A.; Kikkawa, J.; Kobayashi, E.; Asakura, D.; Okubo, M.; Yamada, A., Highly Reversible Oxygen-Redox Chemistry at 4.1 V in $Na_{4/7-x}[\square_{1/7}Mn_{6/7}]O_2$ (□: Mn Vacancy). **2018**, *8*, 1800409.
21. Funke, K.; Roling, B.; Lange, M., Dynamics of Mobile Ions in Crystals, Glasses and Melts. *Solid State Ionics* **1998**, *105*, 195-208.
22. Kobayashi, W.; Yanagita, A.; Akaba, T.; Shimono, T.; Tanabe, D.; Moritomo, Y., Thermal Expansion in Layered $Na_xMO_2$. *Scientific Reports* **2018**, *8*, 3988.
23. Baran, V.; Dolotko, O.; Mühlbauer, M. J.; Senyshyn, A.; Ehrenberg, H., Thermal Structural Behavior of Electrodes in Li-Ion Battery Studied in Operando. *Journal of The Electrochemical Society* **2018**, *165*, A1975-A1982.
24. Schneider, C. A.; Rasband, W. S.; Eliceiri, K. W., Nih Image to Imagej: 25 Years of Image Analysis. *Nat Methods* **2012**, *9*, 671-5.
25. Gerhardt, R., Impedance and Dielectric Spectroscopy Revisited: Distinguishing Localized Relaxation from Long-Range Conductivity. *Journal of Physics and Chemistry of Solids* **1994**, *55*, 1491-1506.
26. Cole, K. S.; Cole, R. H., Dispersion and Absorption in Dielectrics I. Alternating Current Characteristics. *The Journal of Chemical Physics* **1941**, *9*, 341-351.
27. Reddy, M. A.; Fichtner, M., 20 - Ionic Conductivity of Nanocrystalline Metal Fluorides. In *Photonic and Electronic Properties of Fluoride Materials*, Tressaud, A.; Poeppelmeier, K., Eds. Elsevier: Boston, 2016; pp 449-463.
28. Werner, J., Origin of Curved Arrhenius Plots for the Conductivity of Polycrystalline Semiconductors. *Solid State Phenomena* **1994**, *37-38*, 213-218.
29. Briant, J. L., Ionic Conductivity in Lithium and Lithium-Sodium Beta Alumina. *Journal of The Electrochemical Society* **1981**, *128*, 1830.
30. Bhowmik, R. N.; Lone, A. G., Dielectric Properties of $\alpha$-$Fe_{1.6}Ga_{0.4}O_3$ Oxide: A Promising Magneto-Electric Material. *Journal of Alloys and Compounds* **2016**, *680*, 31-42.

31. Funke, K., Ion Transport in Fast Ion Conductors — Spectra and Models. *Solid State Ionics* **1997**, *94*, 27-33.
32. Ravi, M.; Pavani, Y.; Bhavani, S.; Sharma, A. K.; Narasimha Rao, V. V. R., Investigations on Structural and Electrical Properties of $KClO_4$ Complexed Pvp Polymer Electrolyte Films. *International Journal of Polymeric Materials and Polymeric Biomaterials* **2012**, *61*, 309-322.
33. Muthuselvam, I. P.; Bhowmik, R. N., Connectivity between Electrical Conduction and Thermally Activated Grain Size Evolution in Ho-Doped $CoFe_2O_4$ Ferrite. *Journal of Physics D: Applied Physics* **2010**, *43*, 465002.
34. Coşkun, M.; Polat, Ö.; Coşkun, F. M.; Durmuş, Z.; Çağlar, M.; Türüt, A., The Electrical Modulus and Other Dielectric Properties by the Impedance Spectroscopy of $LaCrO_3$ and $LaCr_{0.90}Ir_{0.10}O_3$ Perovskites. *RSC Advances* **2018**, *8*, 4634-4648.
35. Dam, T.; Jena, S. S.; Pradhan, D. K., Coupled Ion Conduction Mechanism and Dielectric Relaxation Phenomenon in $PeO_{20}$–$LiCF_3SO_3$-Based Ion Conducting Polymer Nanocomposite Electrolytes. *The Journal of Physical Chemistry C* **2018**, *122*, 4133-4143.
36. Singh, D. N.; Sinha, T. P.; Mahato, D. K., Electric Modulus, Scaling and Ac Conductivity of $La_2CuMnO_6$ Double Perovskite. *Journal of Alloys and Compounds* **2017**, *729*, 1226-1233.
37. Utpalla, P.; Sharma, S. K.; Sudarshan, K.; Deshpande, S. K.; Sahu, M.; Pujari, P. K., Investigating the Correlation of Segmental Dynamics, Free Volume Characteristics, and Ionic Conductivity in Poly(Ethylene Oxide)-Based Electrolyte: A Broadband Dielectric and Positron Annihilation Spectroscopy Study. *The Journal of Physical Chemistry C* **2020**, *124*, 4489-4501.
38. Taher, Y. B.; Oueslati, A.; Khirouni, K.; Gargouri, M., Dielectric Spectroscopy and Modulus Study of Aluminum Diphosphate $AgAlP_2O_7$. *Journal of Cluster Science* **2015**, *26*, 1655-1669.
39. Hassib, H.; Abdel Razik, A., Dielectric Properties and Ac Conduction Mechanism for 5,7-Dihydroxy-6-Formyl-2-Methylbenzo-Pyran-4-One Bis-Schiff Base. *Solid State Communications* **2008**, *147*, 345-349.
40. Liu, J.; Duan, C.-G.; Yin, W.-G.; Mei, W. N.; Smith, R. W.; Hardy, J. R., Dielectric Permittivity and Electric Modulus in $Bi_2Ti_4O_{11}$. *The Journal of Chemical Physics* **2003**, *119*, 2812-2819.
41. Angell, C. A., Dynamic Processes in Ionic Glasses. *Chemical Reviews* **1990**, *90*, 523-542.
42. Scaife, B. K. P., A New Method of Analysing Dielectric Measurements. *Proceedings of the Physical Society* **1963**, *81*, 124-129.
43. Austin, I. G.; Mott, N. F., Polarons in Crystalline and Non-Crystalline Materials. *Advances in Physics* **1969**, *18*, 41-102.
44. Pollak, M., On the Frequency Dependence of Conductivity in Amorphous Solids. *Philosophical Magazine* **1971**, *23*, 519.
45. Pollak, M.; Pike, G. E., Ac Conductivity of Glasses. *Physical Review Letters* **1972**, *28*, 1449-1451.
46. Long, A. R., Frequency-Dependent Loss in Amorphous Semiconductors. *Advances in Physics* **1982**, *31*, 553-637.
47. Elliott, S. R., A.C. Conduction in Amorphous Chalcogenide and Pnictide Semiconductors. *Advances in Physics* **1987**, *36*, 135-217.
48. Elliott, S. R., Frequency-Dependent Conductivity in Ionically and Electronically Conducting Amorphous Solids. *Solid State Ionics* **1994**, *70-71*, 27-40.
49. Liu, J.; Yin, L.; Wu, L.; Bai, J.; Bak, S.-M.; Yu, X.; Zhu, Y.; Yang, X.-Q.; Khalifah, P. G., Quantification of Honeycomb Number-Type Stacking Faults: Application to Na3ni2bio6 Cathodes for Na-Ion Batteries. *Inorganic Chemistry* **2016**, *55*, 8478-8492.

50. Liu, J.; Hou, M.; Yi, J.; Guo, S.; Wang, C.; Xia, Y., Improving the Electrochemical Performance of Layered Lithium-Rich Transition-Metal Oxides by Controlling the Structural Defects. *Energy & Environmental Science* **2014**, *7*, 705-714.
51. Lu, Z.; Dahn, J. J. C. o. m., Effects of Stacking Fault Defects on the X-Ray Diffraction Patterns of $T_2$, $O_2$, and $O_6$ Structure $Li_{2/3}[Co_XNi_{1/3-X}Mn_{2/3}]O_2$. **2001**, *13*, 2078-2083.
52. Shunmugasundaram, R.; Arumugam, R. S.; Dahn, J. J. J. o. T. E. S., A Study of Stacking Faults and Superlattice Ordering in Some Li-Rich Layered Transition Metal Oxide Positive Electrode Materials. **2016**, *163*, A1394.
53. Liu, J.; Yin, L.; Wu, L.; Bai, J.; Bak, S.-M.; Yu, X.; Zhu, Y.; Yang, X.-Q.; Khalifah, P. G. J. I. c., Quantification of Honeycomb Number-Type Stacking Faults: Application to $Na_3Ni_2BiO_6$ Cathodes for Na-Ion Batteries. **2016**, *55*, 8478-8492.
54. Han, M.; Liu, Z.; Shen, X.; Yang, L.; Shen, X.; Zhang, Q.; Liu, X.; Wang, J.; Lin, H. J.; Chen, C. T. J. A. E. M., Stacking Faults Hinder Lithium Insertion in Li2ruo3. **2020**, *10*, 2002631.
55. Wu, S.; Li, W.; Lin, M.; Burlingame, Q.; Chen, Q.; Payzant, A.; Xiao, K.; Zhang, Q. M., Aromatic Polythiourea Dielectrics with Ultrahigh Breakdown Field Strength, Low Dielectric Loss, and High Electric Energy Density. *Advanced Materials* **2013**, *25*, 1734-1738.
56. Kikkawa, S.; Yasuda, F.; Koizumi, M., Ionic Conductivities of $Na_2Ti_3O_7$, $K_2Ti_4O_9$ and Their Related Materials. *Materials Research Bulletin* **1985**, *20*, 1221-1227.
57. Chatterjee, S.; Mahapatra, P.; Choudhary, R. N.; Thakur, A., Complex Impedance Studies of Sodium Pyrotungstate – $Na_2W_2O_7$. *physica status solidi (a)* **2004**, *201*, 588-595.
58. Molenda, J.; Stokłosa, A.; Than, D., Relation between Ionic and Electronic Defects of $Na_{0.7}MnO_2$ Bronze and Its Electrochemical Properties. *Solid State Ionics* **1987**, *24*, 33-38.
59. Blesa, M. C.; Moran, E.; León, C.; Santamaria, J.; Tornero, J. D.; Menéndez, N., α-$NaFeO_2$: Ionic Conductivity and Sodium Extraction. *Solid State Ionics* **1999**, *126*, 81-87.
60. Tuller, H. L.; Moon, P. K., Fast Ion Conductors: Future Trends. *Materials Science and Engineering: B* **1988**, *1*, 171-191.
61. Bera, A. K.; Yusuf, S. M., Temperature-Dependent Na-Ion Conduction and Its Pathways in the Crystal Structure of the Layered Battery Material $Na_2Ni_2TeO_6$. *The Journal of Physical Chemistry C* **2020**, *124*, 4421-4429.
62. Li, Y.; Deng, Z.; Peng, J.; Gu, J.; Chen, E.; Yu, Y.; Wu, J.; Li, X.; Luo, J.; Huang, Y.; Xu, Y.; Gao, Z.; Fang, C.; Zhu, J.; Li, Q.; Han, J.; Huang, Y., New P2-Type Honeycomb-Layered Sodium-Ion Conductor: $Na_2Mg_2TeO_6$. *ACS Applied Materials & Interfaces* **2018**, *10*, 15760-15766.
63. Deng, Z.; Gu, J.; Li, Y.; Li, S.; Peng, J.; Li, X.; Luo, J.; Huang, Y.; Fang, C.; Li, Q.; Han, J.; Huang, Y.; Zhao, Y., Ca-Doped $Na_2Zn_2TeO_6$ Layered Sodium Conductor for All-Solid-State Sodium-Ion Batteries. *Electrochimica Acta* **2019**, *298*, 121-126.
64. Zaafouri, A.; Megdiche, M.; Gargouri, M., Ac Conductivity and Dielectric Behavior in Lithium and Sodium Diphosphate $LiNa_3P_2O_7$. *Journal of Alloys and Compounds* **2014**, *584*, 152-158.
65. Ben Rhaiem, A.; Hlel, F.; Guidara, K.; Gargouri, M., Electrical Conductivity and Dielectric Analysis of $AgNaZnP_2O_7$ Compound. *Journal of Alloys and Compounds* **2009**, *485*, 718-723.
66. Mahamoud, H.; Louati, B.; Hlel, F.; Guidara, K., Conductivity and Dielectric Studies on $(Na_{0\cdot4}Ag_{0\cdot6})_2PbP_2O_7$ compound. *Bulletin of Materials Science* **2011**, *34*, 1069-1075.
67. Yaroslavtsev, A. B.; Stenina, I. A., Complex Phosphates with the Nasicon Structure $(M_xA_2(PO_4)_3$. *Russian Journal of Inorganic Chemistry* **2006**, *51*, S97-S116.

68. Smaha, R. W.; Roudebush, J. H.; Herb, J. T.; Seibel, E. M.; Krizan, J. W.; Fox, G. M.; Huang, Q.; Arnold, C. B.; Cava, R. J., Tuning Sodium Ion Conductivity in the Layered Honeycomb Oxide $Na_{3-X}Sn_{2-X}Sb_xNaO_6$. *Inorganic Chemistry* **2015**, *54*, 7985-7991.
69. Liu, J.; Chang, D.; Whitfield, P.; Janssen, Y.; Yu, X.; Zhou, Y.; Bai, J.; Ko, J.; Nam, K. W.; Wu, L.; Zhu, Y.; Feygenson, M.; Amatucci, G.; Van der Ven, A.; Yang, X. Q.; Khalifah, P., Ionic Conduction in Cubic $Na_3TiP_3O_9N$, a Secondary Na-Ion Battery Cathode with Extremely Low Volume Change. *Chemistry of Materials* **2014**, *26*, 3295-3305.
70. Shanmugam, R.; Lai, W., Study of Transport Properties and Interfacial Kinetics of $Na_{2/3}[Ni_{1/3}Mn_xTi_{2/3-X}]O_2$(X = 0,1/3) as Electrodes for Na-Ion Batteries. *Journal of The Electrochemical Society* **2014**, *162*, A8-A14.

**TABLE 1.** The values of activation energies Ea (in eV) at selected frequencies as determined from the temperature-dependent ac conductivity data.

| **Frequency (Hz)** | $E_a$ **(eV)** (383-518 K) | $E_a$ **(eV)** (518-713 K) |
|---|---|---|
| 1 | 0.156 ± 0.007 | 0.383 ± 0.006 |
| 10 | 0.156 ± 0.006 | 0.383 ± 0.006 |
| $10^2$ | 0.156 ± 0.006 | 0.378 ± 0.006 |
| $10^3$ | 0.155 ± 0.007 | 0.374 ± 0.006 |
| $10^4$ | 0.145 ± 0.007 | 0.371 ± 0.006 |
| $10^5$ | 0.107 ± 0.007 | 0.363 ± 0.007 |

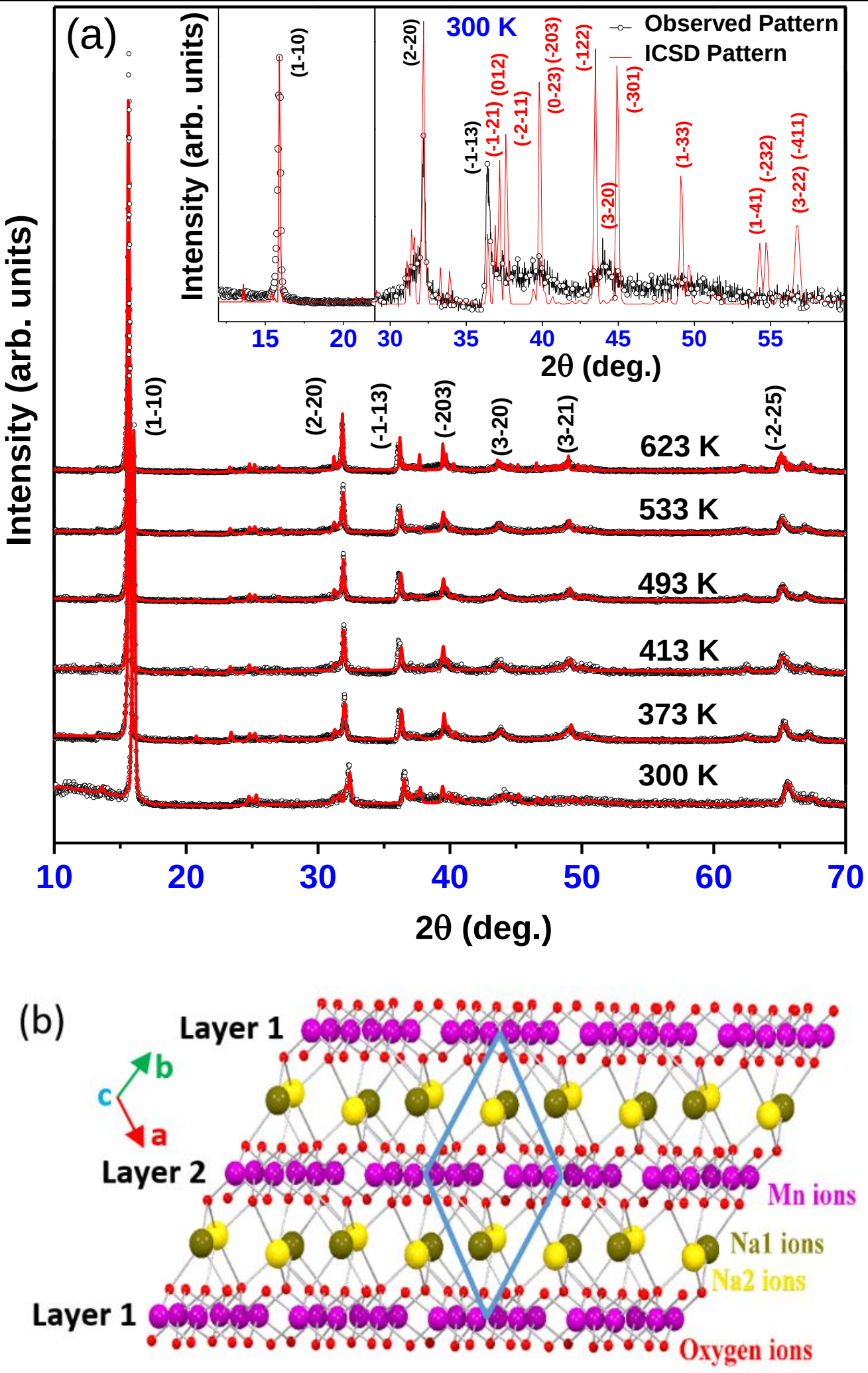

**Figure 1.** (a) Temperature-dependent x-ray diffraction patterns of $Na_2Mn_3O_7$ for the sample annealed for 4 hours. The symbols represent experimental data points and the solid red lines through the data points are the calculated patterns. The Miller indices for the intensed Bragg peaks are also mentioned. The inset shows a comparison of the observed x-ray diffraction pattern (black line), measured at 300 K, with that for the ideal crystal structure without stacking faults (red line). (b) Layered crystal structure of $Na_2Mn_3O_7$ showing the $Mn_3O_7^{2-}$ layers. The blue box represents the dimension of the unit cell.

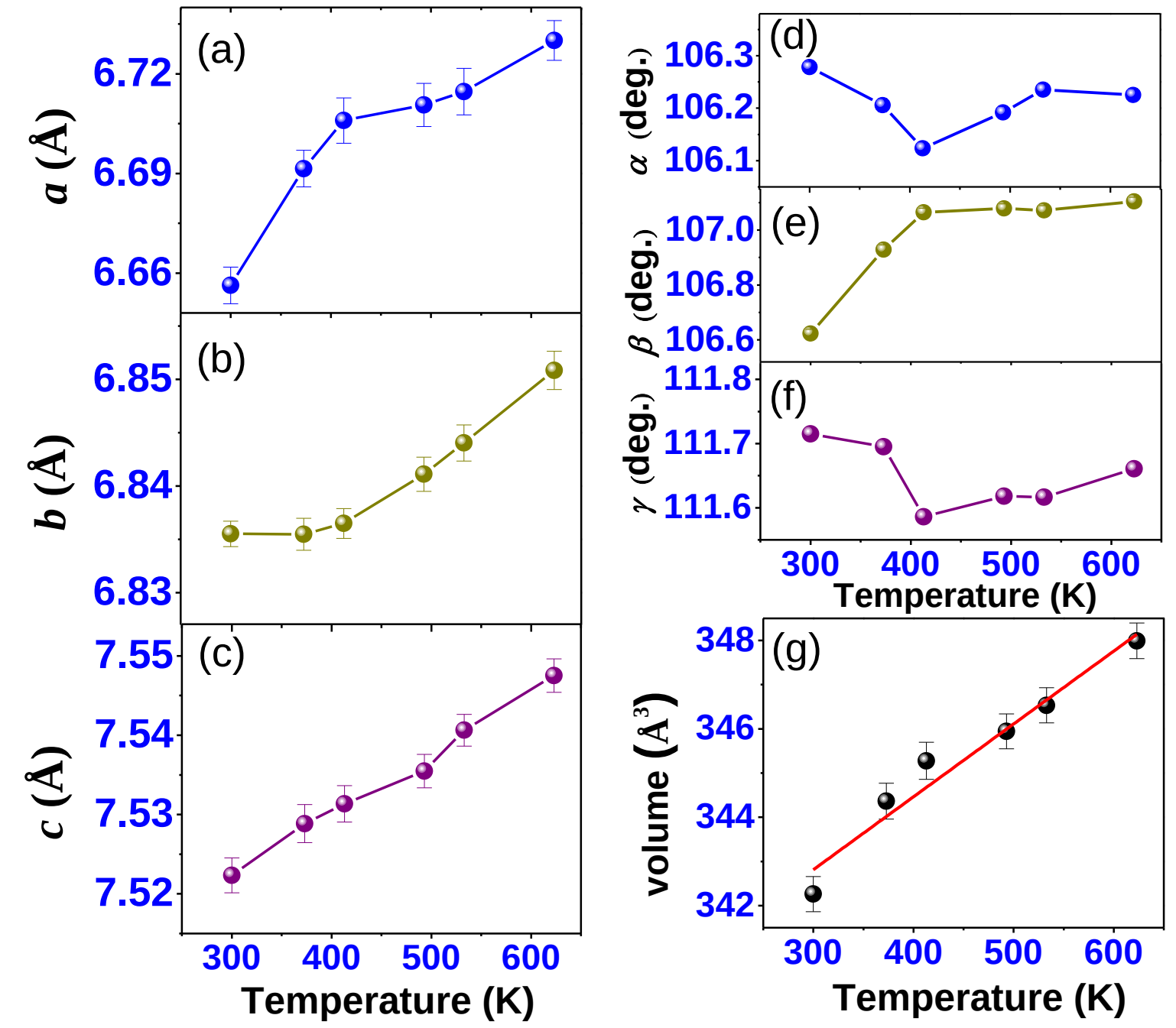

**Figure 2.** (a) - (f) Temperature-dependent lattice parameters *a*, *b*, *c*, *α*, *β*, and *γ* of $Na_2Mn_3O_7$, respectively. The solid lines are guide to the eyes. (g) The variation of the unit cell volume of $Na_2Mn_3O_7$ with temperature. The solid line is a linear fit to the experimental data points.

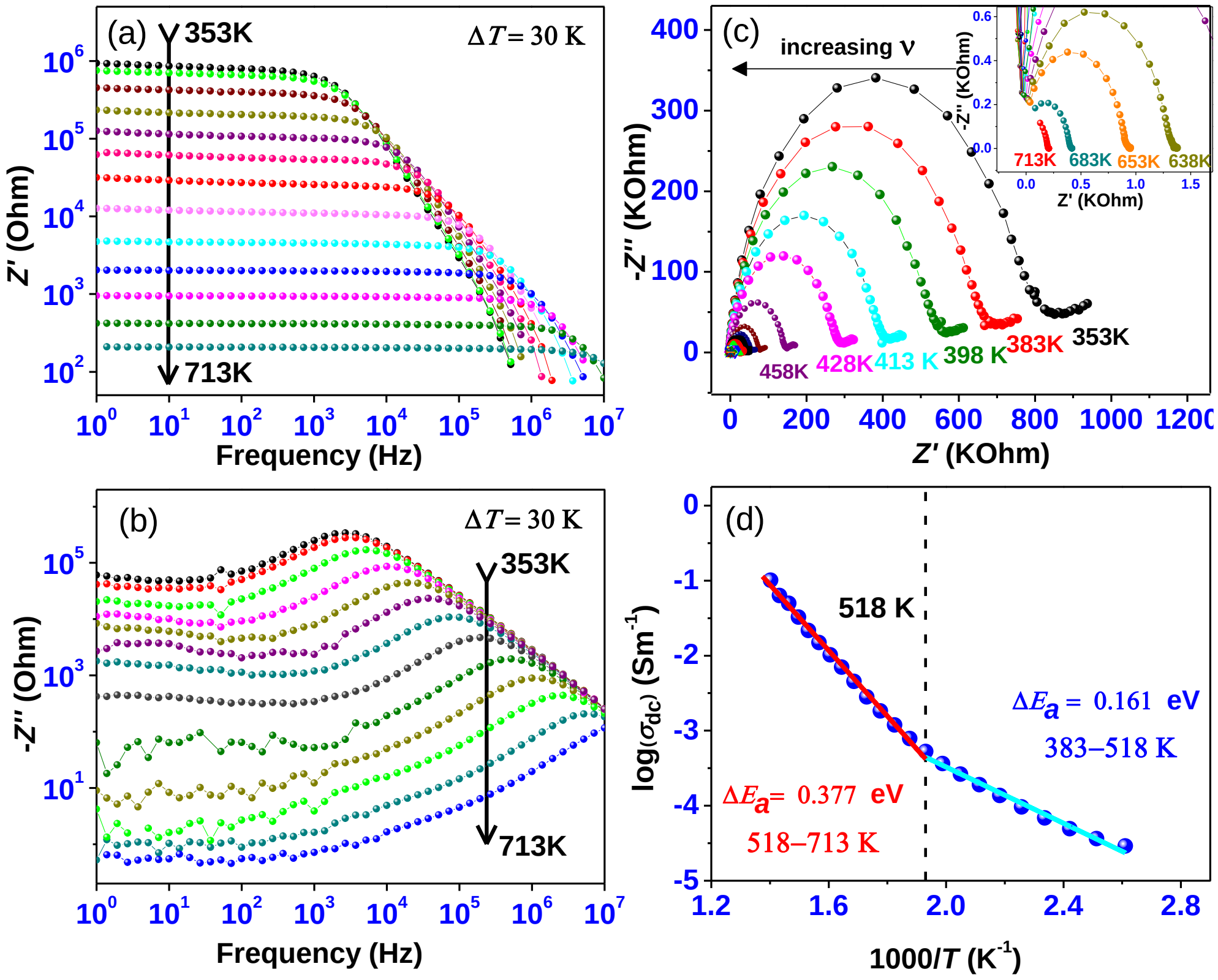

**Figure 3.** (a) Real ($Z'$) and (b) imaginary ($-Z''$) parts of impedance of $Na_2Mn_3O_7$, respectively, as a function of frequency, measured over 353-713 K. (c) The $-Z''$ vs $Z'$ curves (Cole-Cole plot) for $Na_2Mn_3O_7$ at selected temperatures. (d) Temperature dependence of the dc conductivity ($\sigma_{dc}$) for the compound $Na_2Mn_3O_7$.

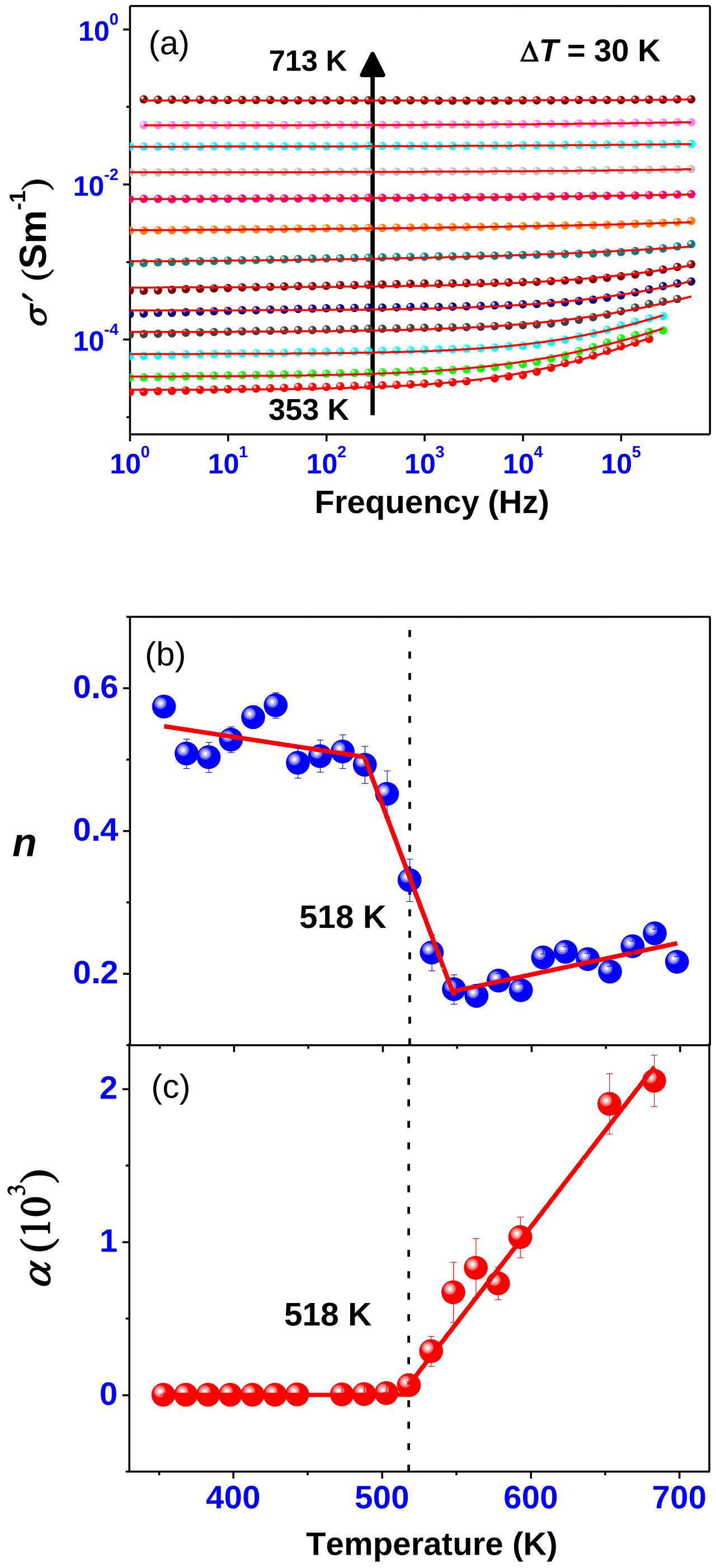

**Figure 4.** (a) Frequency-dependent real part of the ac conductivity ($\sigma'$) of $Na_2Mn_3O_7$ at different temperatures. The solid lines through the data points are the fitted curves as per the Jonscher′s power law (Eq.12). (b) and (c) The temperature dependences of the frequency exponent $n$ and the parameter $\alpha$, respectively.

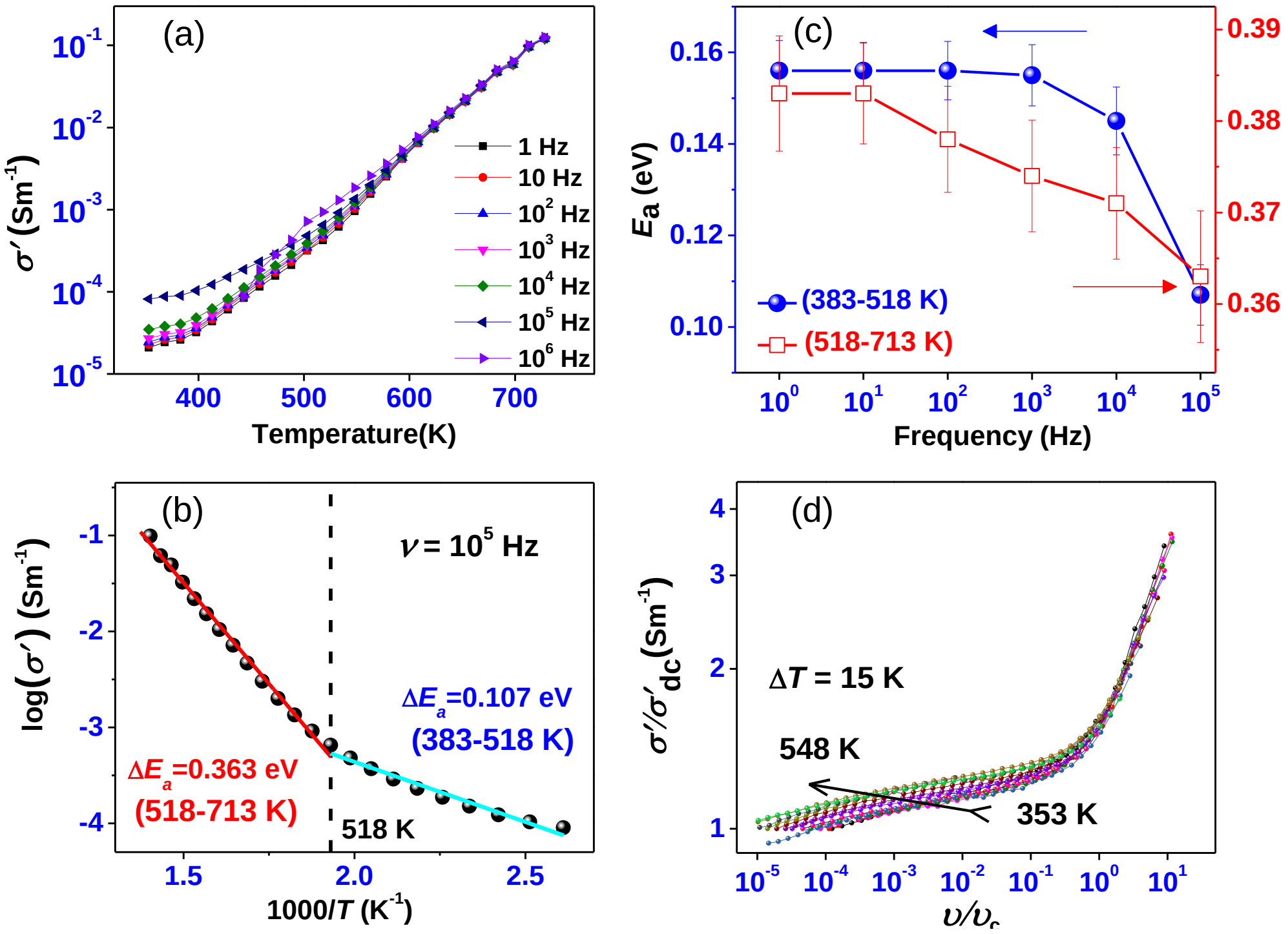

**Figure 5**. (a) Temperature dependence of the ac conductivity ($\sigma'$) of $Na_2Mn_3O_7$ at selected frequencies. (b) The Arrhenius plot of the $\sigma'(T)$ for the frequency $\nu = 10^5$ Hz. (c) The variation of activation energies as a function of frequency. (d) Normalized ac conductivity ($\sigma'/\sigma'_{dc}$) vs normalized frequency ($\nu/\nu_c$) curves at selected temperatures.

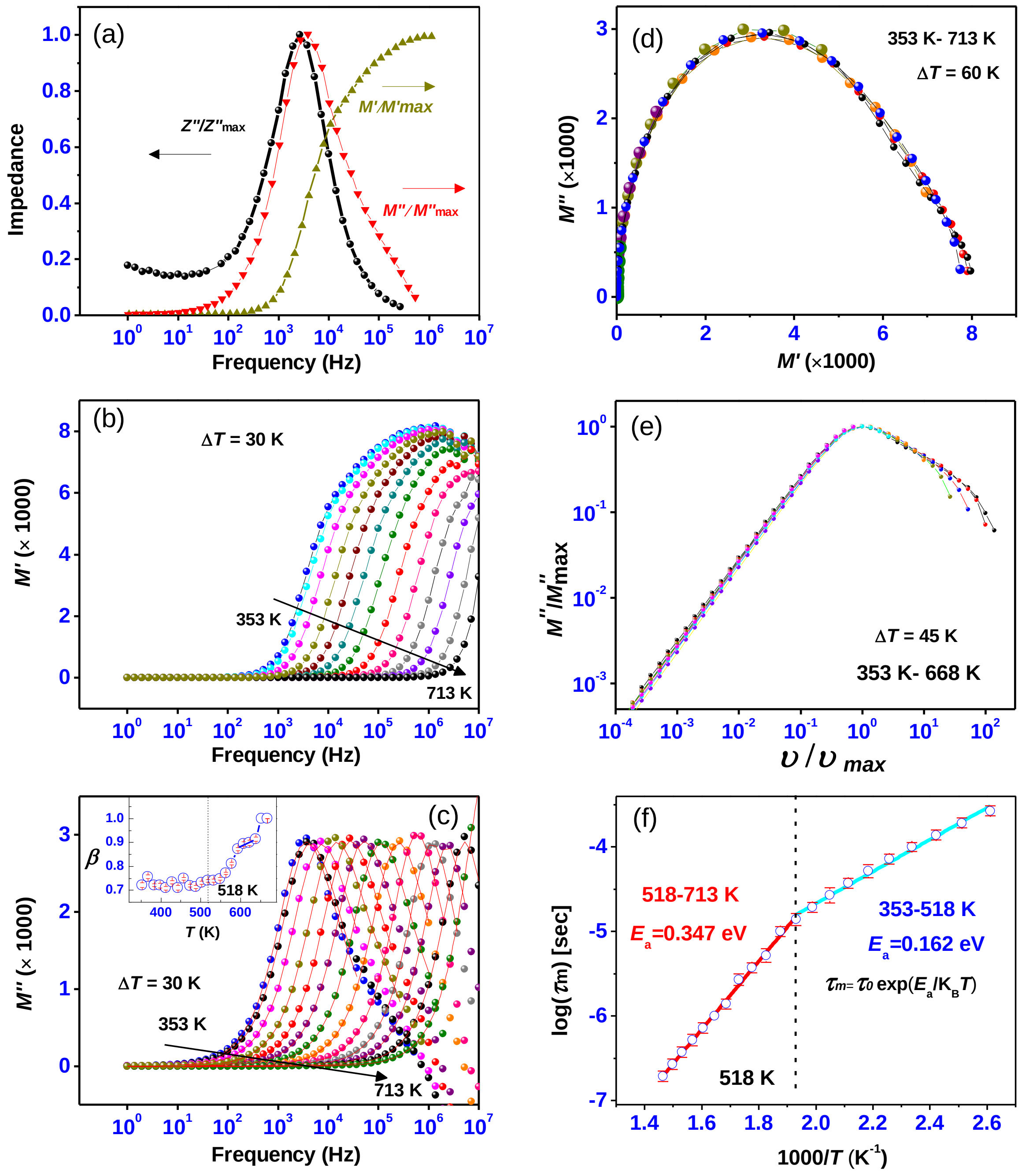

**Figure 6.** (a) A comparative plot between normalized imaginary part of impedance ($Z''/Z''_{max}$) and normalized real and imaginary parts of electrical modulus ($M''/M''_{max}$, $M'/M'_{max}$), measured at 353 K, of $Na_2Mn_3O_7$. (b) and (c) The frequency-dependent real ($M'$) and imaginary ($M''$) parts of the electrical modulus at selected temperatures. The solid curves in (c) are the fitted patterns by Eq. 16. The inset of (c) shows the temperature dependence of the stretched exponential parameter $\beta$. (d) Cole-Cole plot in electrical modulus plane ($M''$ vs $M'$). (e) Universal scaling behaviour of the imaginary part of electrical modulus ($M''/M''_{max}$ vs $\nu/\nu_{max}$ at selected temperatures). (f) Arrhenius plot of the temperature -dependent relaxation time ($\tau_m$).

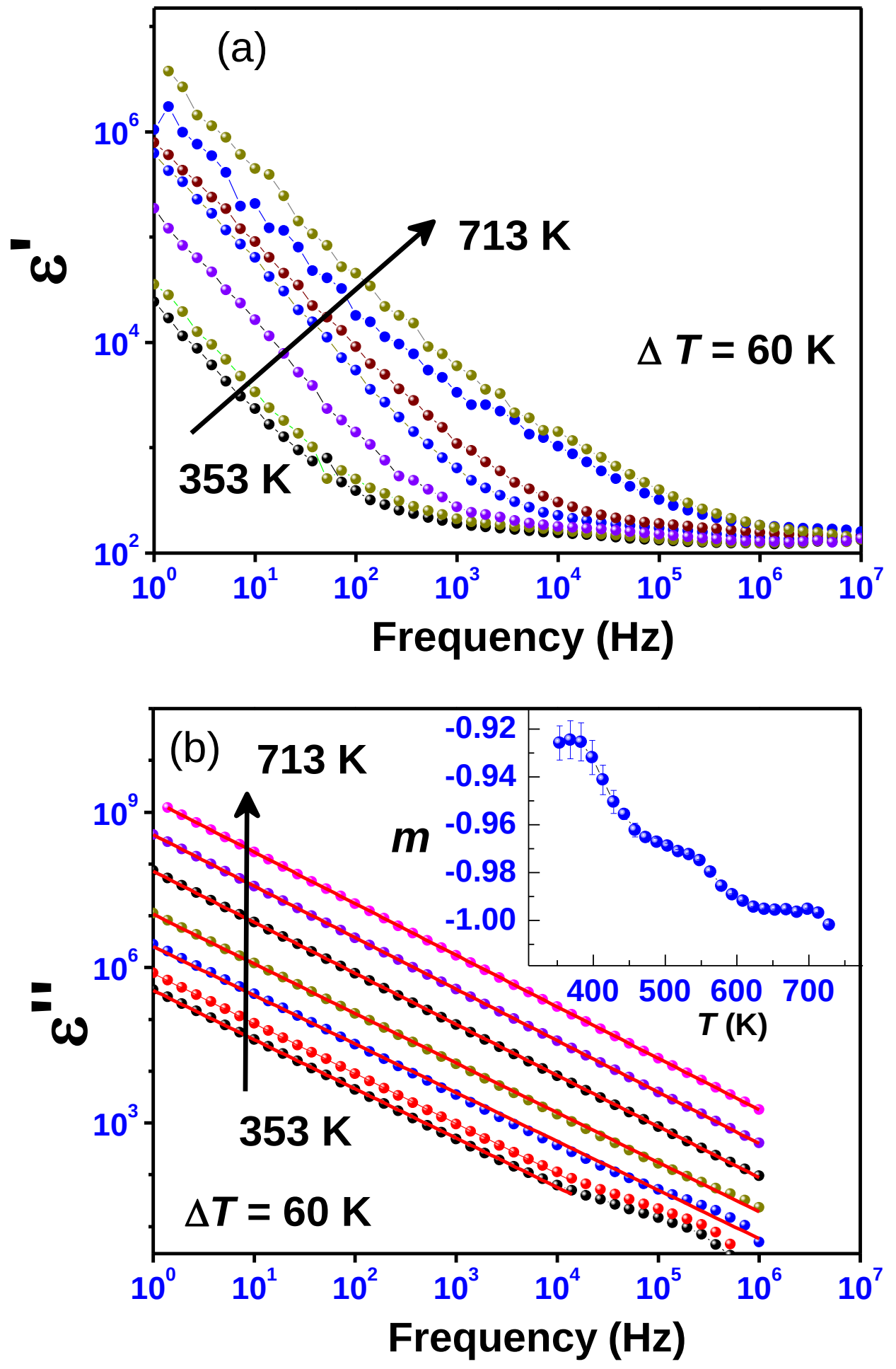

**Figure 7.** (a) and (b) The frequency-dependent real ($\varepsilon'$) and imaginary ($\varepsilon''$) parts of the dielectric constant of $Na_2Mn_3O_7$ at selected temperatures. The solid lines in (b) are the fitted curve as per the Eq. 22. The inset of (b) shows the temperature-dependent exponent $m(T)$ curves.

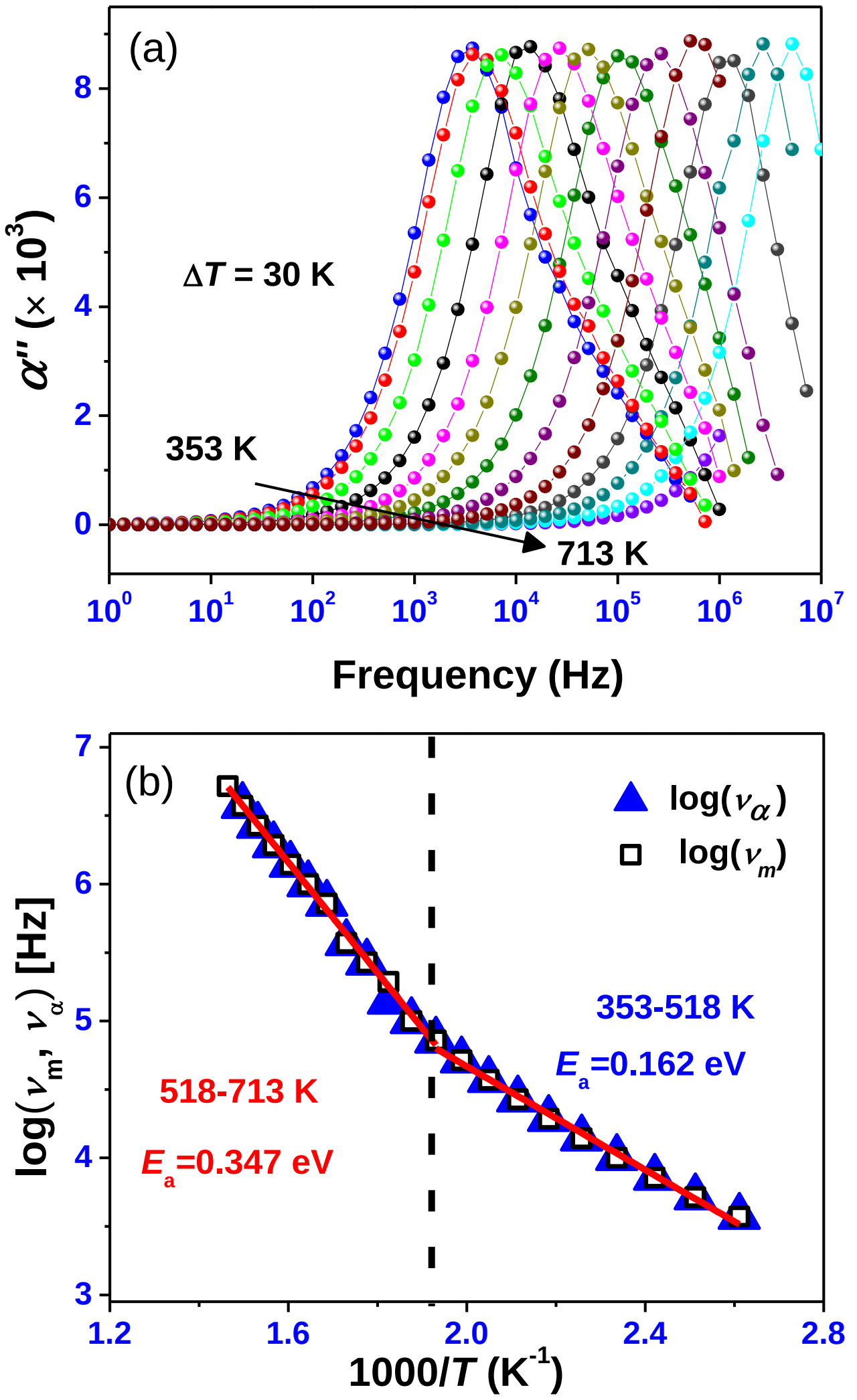

**Figure 8.** (a) Frequency dependent imaginary part of the complex polarizability ($\alpha''$) curves at selected temperatures. (b) Arrhenius plots of the peak frequencies $\nu_\alpha$ and $\nu_m$, obtained from the imaginary parts of the complex polarizability ($\alpha''$) and complex modulus ($M''$) [Fig. 7(c)], respectively.

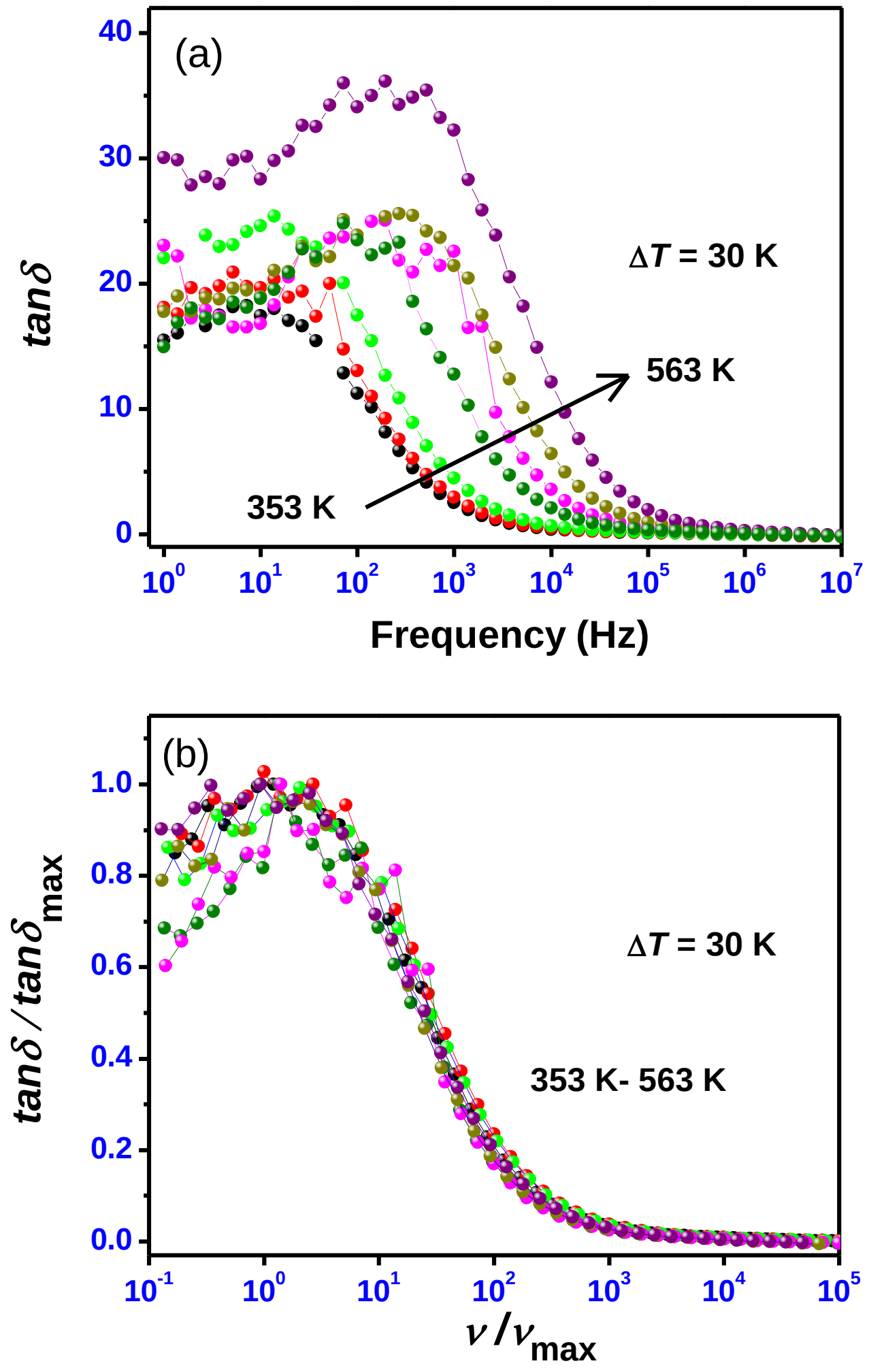

**Figure 9.** (a) The frequency dependence of the dielectric loss factor (*tanδ*) for $Na_2Mn_3O_7$ at different temperatures. (b) Scaling behavior of the loss factor ($tan\delta/tan\delta_{max}$ vs $\nu/\nu_{max}$).

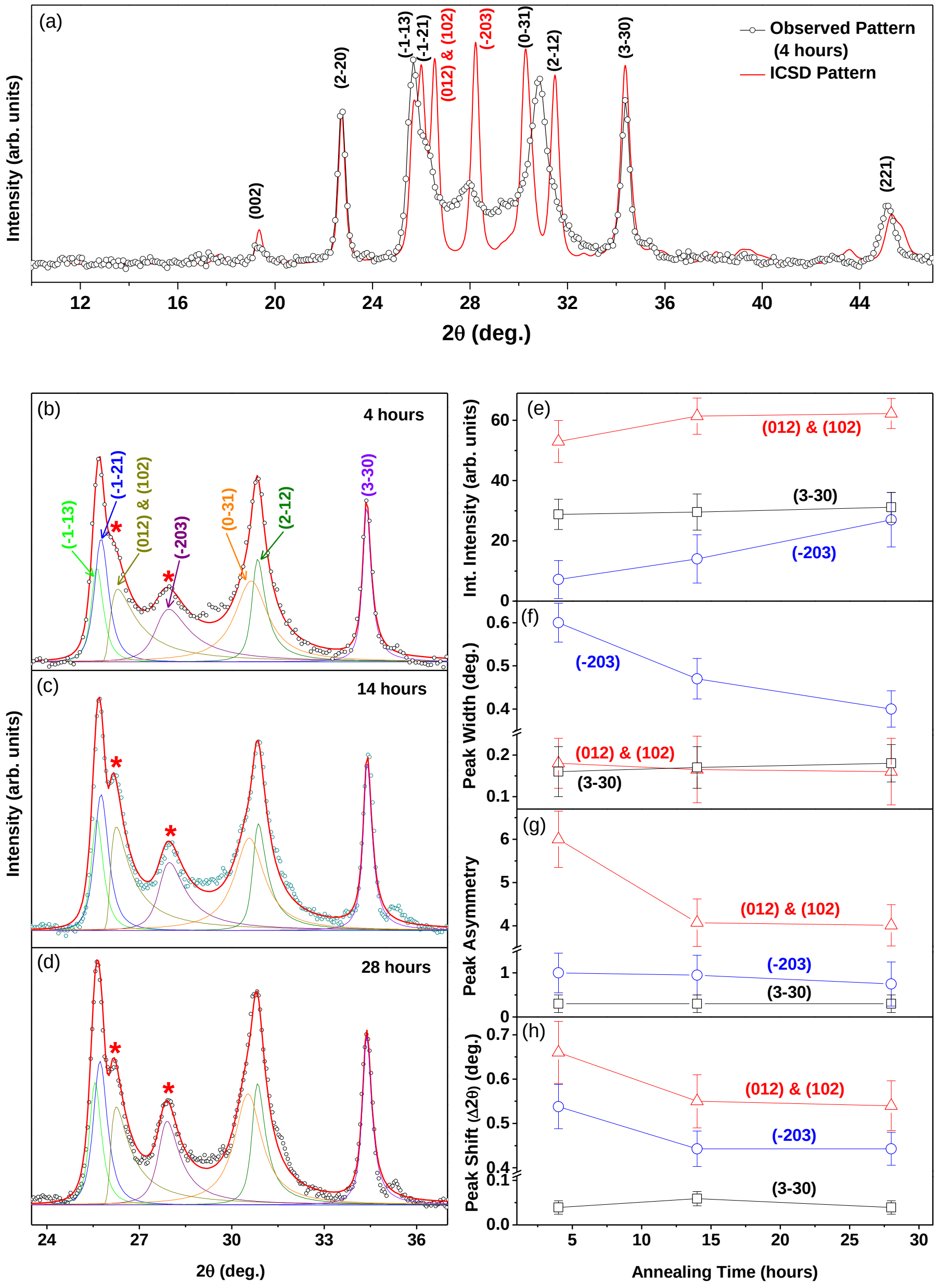

**Figure 10.** Room temperature neutron diffraction patterns for $Na_2Mn_3O_7$ synthesized with different annealing times. (a) A comparison of the observed pattern (for 4 hours annealed sample) with that for the ideal crystal structure without stacking faults (ICSD_collcode5665). (b-d) Experimental neutron diffraction patterns (symbols) of the samples synthesized with the annealing times 4, 14 and 28 hours, respectively. The solid lines are calculated peak profiles with the asymmetric PearsonIV function. The Bragg peaks (012), (102) and (-203) (marked with asterisks) show substantial changes with increasing annealing time. (e-h) The variations of the integrated intensity, peak width, asymmetric parameter and peak shift for (012), (102), (-203) Bragg peaks, derived from the neutron diffraction patterns, with annealing time, respectively.

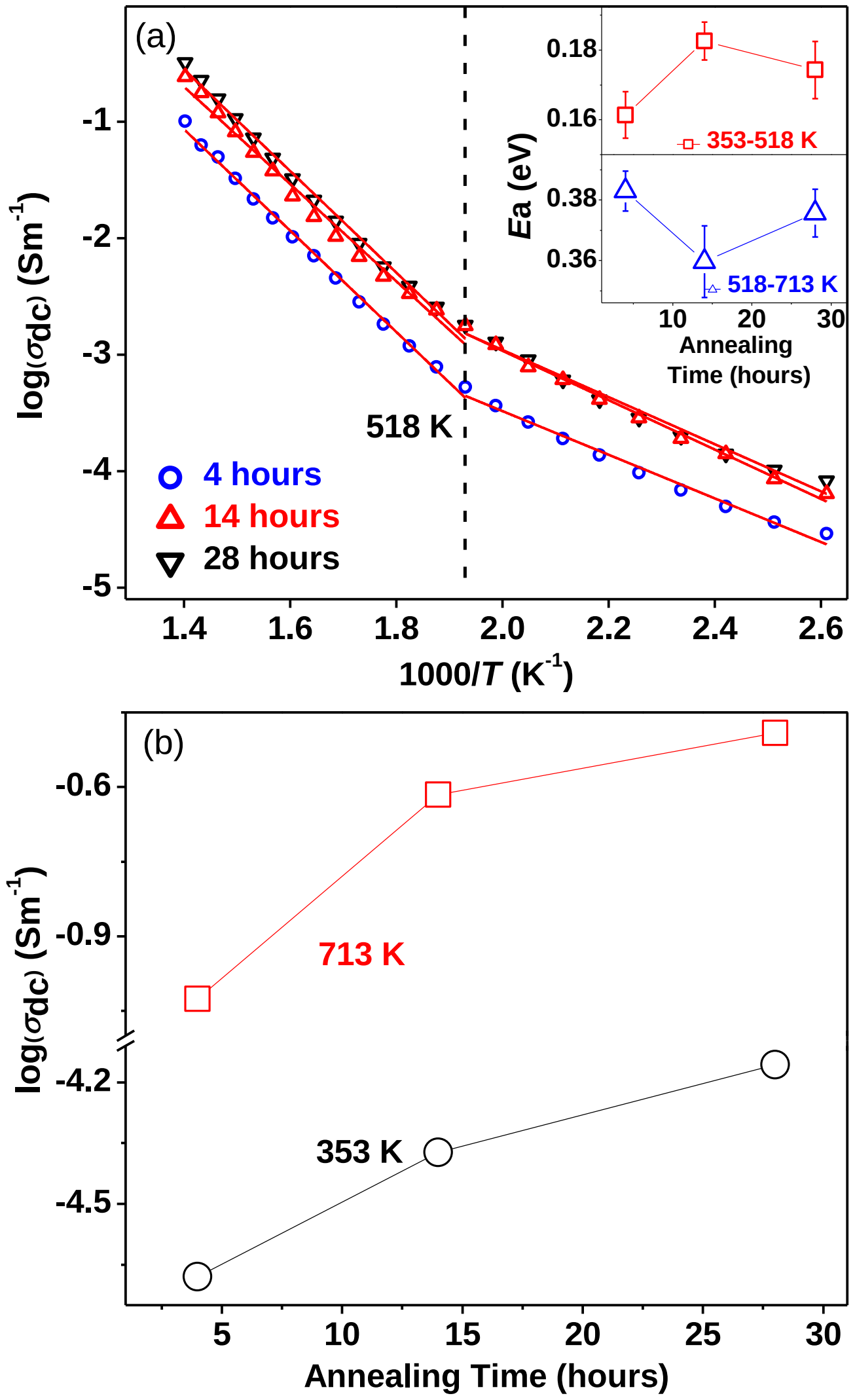

**Figure 11.** (a) Variation of dc ionic conductivity of $Na_2Mn_3O_7$ with different annealing time. The red line is fitted with Arrhenius equation for two temperature ranges 353 K-518 K and 518 K-713 K. The inset shows the variation of activation energies with the annealing time. (b) The variation of the dc ionic conductivities with the annealing time at selected temperatures 353 and 713 K.

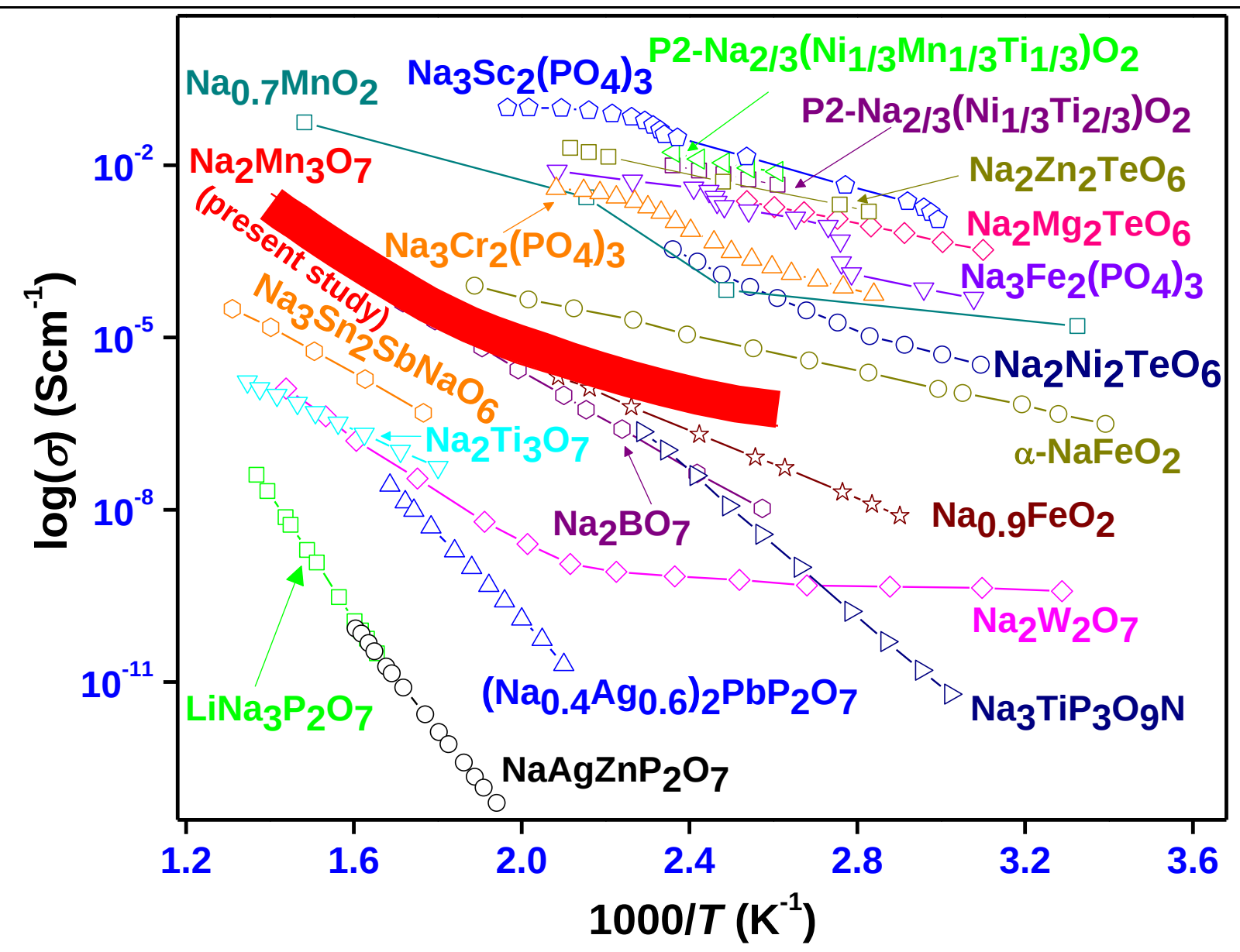

**Figure 12.** Temperature-dependent ionic conductivity of $Na_2Mn_3O_7$ (the width of the line represents the conductivity values of $Na_2Mn_3O_7$ synthesized at different annealing time 4, 14, and 28 h) compared with other related Na-ion based layered compounds: $Na_2Ti_3O_7$,[56] $Na_2W_2O_7$,[57] $Na_{0.7}MnO_2$,[58] α-$NaFeO_2$,[59] $Na_2BO_7$,[60] $Na_2Ni_2TeO_6$,[61] $Na_2Mg_2TeO_6$,[62] $Na_2Zn_2TeO_6$,[63] $LiNa_3P_2O_7$,[64] $NaAgZnP_2O_7$,[65] Na0.4$Ag_{0.62}PbP_2O_7$,[66] $Na_3Fe_2(PO_4)_3$,[67] $Na_3Cr_2(PO_4)_3$,[67] $Na_3Sc_2(PO_4)_3$,[67] $Na_3Sn_2SbNaO_6$,[68] $Na_3TiP_3O_9N$,[69] P2-$Na_{2/3}Ni_{1/3}Ti_{2/3}O_2$,[70] and P2-$Na_{2/3}Ni_{1/3}Mn_{1/3}Ti_{1/3}O_2$.[70]

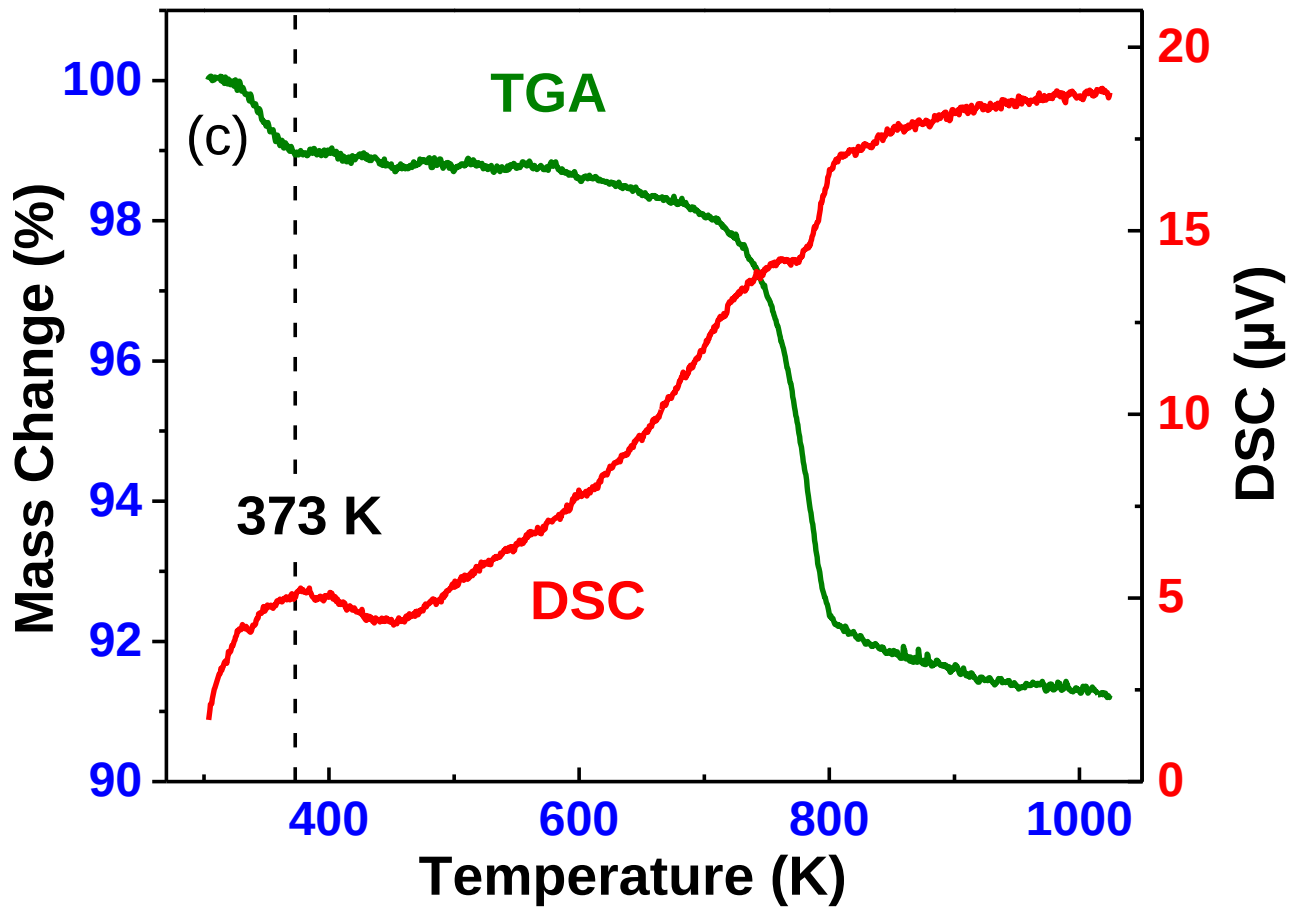

**Figure S1.** The temperature-dependent TGA and DSC curves for $Na_2Mn_3O_7$.

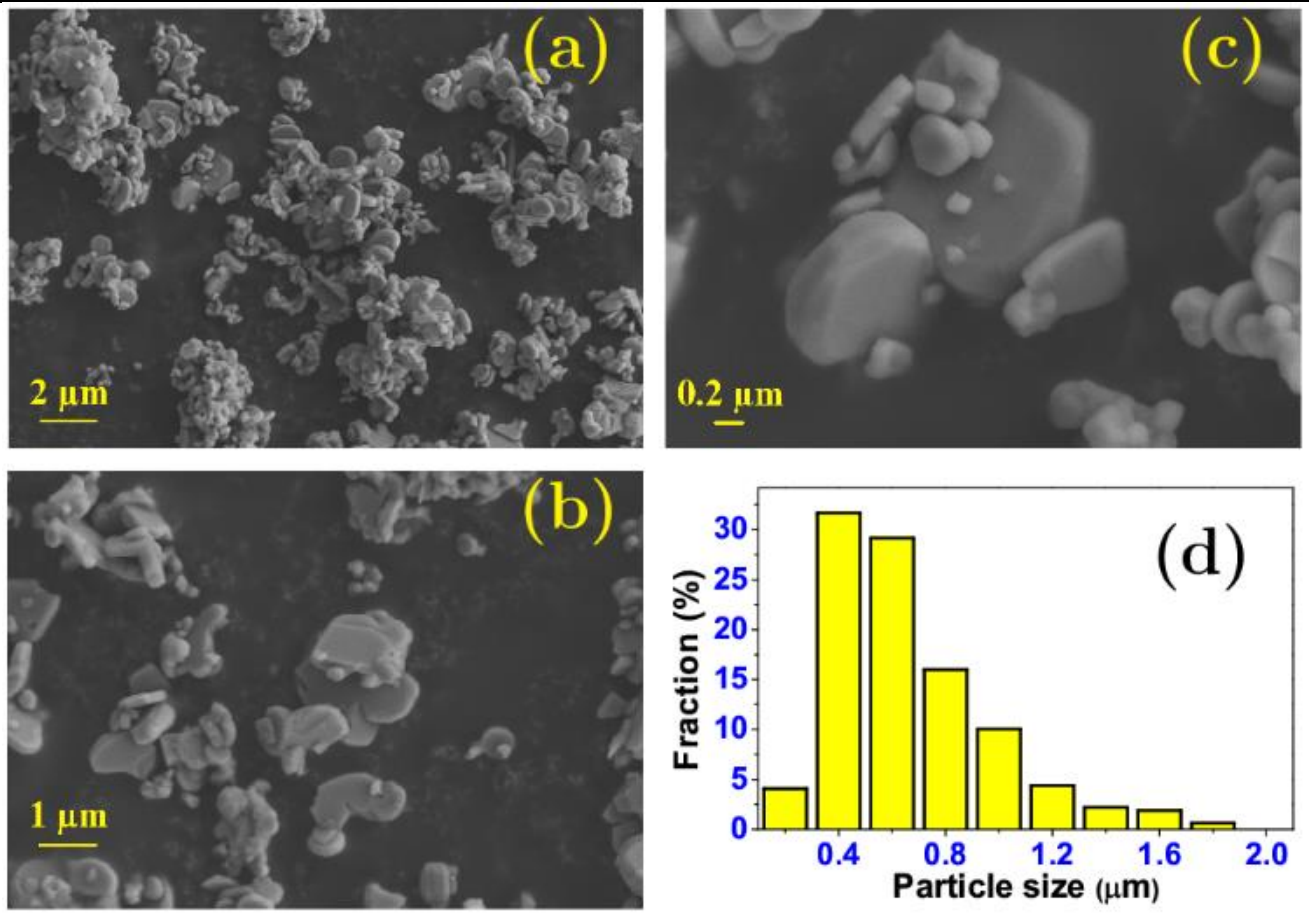

**Figure S2.** (a), (b) and (c) High resolution scanning electron microscope images of the polycrystalline samples of $Na_2Mn_3O_7$ measured under different magnifications of 5000×, 10000× and 25000×, respectively. (d) The particle size distribution of the polycrystalline samples of $Na_2Mn_3O_7$ as derived from the SEM images.